\documentclass{aa}

\newcommand{\target}{5BZQ\,J0808+4950}
\newcommand{\targetgamma}{4FGL\,J0808.5+4950}
\newcommand{\gbneutrino}{GB6\,J1040+0617}
\newcommand{\txs}{TXS\,0506+056}
\newcommand{\pksseven}{PKS\,0735+178}
\newcommand{\pksfifteen}{PKS\,1502+106}
\newcommand{\icevent}{IC110929A}

\newcommand{\MgII}{Mg~\MakeUppercase{\romannumeral2}\,}

\newcommand{\origin}{\textsc{Paper~I}}
\newcommand{\originII}{\textsc{Paper~II}}
\newcommand{\paperPHone}{\textsc{Paper~PI}} 
\newcommand{\paperNUone}{\textsc{Paper~NI}} 

\usepackage{hyperref}
\usepackage{amsmath}
\usepackage{multirow}
\usepackage{url}
\usepackage{adjustbox}
\usepackage{xcolor}
\usepackage{ulem}
\usepackage{lineno}
\begin{document}
\linenumbers
\title{A lepto-hadronic perspective on the properties of the candidate neutrino-emitter blazar 5BZQ\,J0808+4950}

\author{M. Boughelilba\inst{1} \corrauth{margot.boughelilba@desy.com} \and L. Pfeiffer\inst{2} \and J.M. Sanchez Zaballa\inst{1,2} \and A. Azzollini\inst{1,2} \and A. Bremer\inst{2} \and S. Buson\inst{1,2} \and M. Lincetto\inst{1} }

\institute{Deutsches Elektronen-Synchrotron DESY, Platanenallee 6, 15738 Zeuthen, Germany \and Julius-Maximilians-Universit{\"a}t W{\"u}rzburg, Fakult{\"a}t f{\"u}r Physik und Astronomie, Institut f{\"u}r Theoretische Physik und Astrophysik, Lehrstuhl f{\"u}r Astronomie, Emil-Fischer-Str. 31, D-97074 W{\"u}rzburg, Germany}
  
\date{Submitted 03-11-2026}

\abstract
{The role of active galactic nuclei in the production of high-energy neutrinos has been highlighted by the association of the blazar TXS\,0506+056 and the Seyfert galaxy NGC 1068 with IceCube neutrino events. In particular, blazars, with their powerful jets, are potential sites of neutrino production via (photo-)hadronic interactions.}
{The gamma-ray bright blazar 5BZQ\,J0808+4950 is consistent with a spatial anisotropy in the IceCube neutrino all-sky maps. An individual high-energy neutrino event, IC110929A, was also reported in the direction of this blazar. Although the correlation between the gamma-ray bright blazars and the neutrino production remains inconclusive,
5BZQ\,J0808+4950, with a luminous accretion disk serving as a photon field for the photo-hadronic interactions, offers the opportunity to test 
the relation between neutrino production and the radiative output of the blazar jet.}
{We compile available multi-wavelength data and analyze the gamma-ray light curve of 5BZQ\,J0808+4950, leading us to define six contemporaneous epochs to model the source's spectral energy distribution and its evolution over time. We use a lepto-hadronic, time-dependent framework to model 5BZQ\,J0808+4950, taking into account external photon fields.}
{We find that the solutions corresponding to the different states the source exhibits are explained mostly by a change in the particles' injection properties. The correlation between the predicted X-ray and the neutrino luminosity in this source is stronger than the correlation between gamma-ray and neutrino luminosity. We predict that the rate of neutrinos detectable by IceCube for this blazar ranges from $\sim 10^{-3}$ to $\sim10^{-1}$ neutrinos per epoch. Assuming Poisson statistics, the predicted neutrino number contemporaneous with the interval in which the high-energy neutrino event IC110929A was detected, is compatible with the detection of one neutrino.}
{}

\keywords{}
\authorrunning{M. Boughelilba et al.}
\titlerunning{Lepto-hadronic perspective on \target}
\maketitle
\nolinenumbers

\section{INTRODUCTION} 
\label{sec:intro}

Active galactic nuclei (AGN), powered by accretion onto a central supermassive black hole, have since long been considered to be good candidates for the acceleration of (ultra-) high-energy cosmic rays \citep{hillas_origin_1984}. If AGN jets are able to accelerate hadrons to sufficient energy, these particles can interact with photon fields \citep[$p\gamma$ interactions; see][for an application to astrophysical environments]{mucke_photohadronic_1999} and in turn produce neutrinos \citep[see e.g.][]{mannheim_photomeson_1989, stecker_highenergy_1991,stecker_erratum_1992, mannheim_proton_1993, szabo_implications_1994, mannheim_highenergy_1995, mastichiadis_hadronic_1996, protheroe_high_1999, atoyan_highenergy_2001, mucke_proton_2001, atoyan_neutral_2003, mucke_bl_2003, dermer_ultrahighenergy_2009, dermer_variable_2012, dimitrakoudis_timedependent_2012, bottcher_leptonic_2013, halzen_pionic_2013, murase_diffuse_2014, dermer_photopion_2014, mastichiadis_consequences_2016, gao_direct_2017, murase_blazar_2018, oikonomou_highenergy_2019}. Non-jetted AGN are also considered as possible neutrino emitters, invoking hadronic interactions ($pp$ interactions) taking place at the core of the AGN \citep[see e.g.][]
{murase_hidden_2016, inoue_highenergy_2019, inoue_gammaray_2021, kheirandish_highenergy_2021, fiorillo_tev_2024, karavola_neutrino_2025, neronov_neutrino_2025}. In both cases, neutrinos are the smoking-gun signatures of hadronic interactions. 

The current most sensitive detector for high-energy ($\gtrsim$~TeV) astrophysical neutrino searches is the IceCube Neutrino observatory, located at the South Pole \citep{IceCubecollaboration_evidence_2013, aartsen_first_2013, aartsen_observation_2014a, aartsen_combined_2015, aartsen_evidence_2015}. As of today, only a few high-energy neutrino events detected by IceCube are spatially consistent with astrophysical sources that are thought to be capable of producing them. Currently, the strongest astrophysical neutrino signal originates from the nearby Seyfert galaxy NGC 1068 -- a non-jetted, low-luminosity AGN -- \citep{IceCubecollaboration_evidence_2022}. In the jetted AGN class, instead, some of the most promising sources are blazars (jetted AGN with a jet aligned with our line of sight) such as PKS 1424-41 \citep{kadler_coincidence_2016}, \gbneutrino\ \citep{garrappa_investigation_2019}, 3HSP J095507.9+355101 \citep{giommi_3hsp_2020}, \pksfifteen\  \citep{franckowiak_patterns_2020} and \pksseven\  \citep{sahakyan_multimessenger_2023}.

So far, the most significant association between a high-energy neutrino event and a blazar is the one of \txs\ which was found to be spatially coincident with the $290$-TeV neutrino IceCube-170922A \citep{icecubecollaboration_neutrino_2018}, while the blazar was experiencing a gamma-ray flare \citep{tanaka_fermilat_2017}. 
Recently, the detection by the Cubic Kilometer Neutrino Telescope (KM3NeT) of an ultra-high energy (UHE) neutrino \citep[with an estimated energy $\sim 220 \, \rm PeV$,][]{thekm3netcollaboration_observation_2025,thekm3netcollaboration_author_2025} challenged further our understanding of the astrophysical neutrino production. In a recent work \citep{km3netcollaboration_characterizing_2025}, 17 blazars are found to be positionally consistent with the neutrino arrival direction, although no firm conclusion can be drawn from this study.
Even if the contribution of gamma-ray bright blazars to the diffuse neutrino flux measured by IceCube is suggested to be limited to $\sim 27\%$ \citep{aartsen_contribution_2017,abbasi_search_2023}, several studies point out the interest of subclasses of blazars regarding high-energy neutrino correlation (\citealt{giommi_dissecting_2020}, \citealt{hovatta_association_2021}, \citealt{plavin_directional_2021}, \citealt{zhou_search_2021}, \citealt{buson_beginning_2022}, hereafter \origin, \citealt{buson_erratum_2022}, \citealt{padovani_spectra_2022}, \citealt{buson_extragalactic_2023}, hereafter \originII, \citealt{karl_spectra_2023}, \citealt{rodrigues_spectra_2024}, \citealt{moretti_flatspectrum_2025}).

In addition to high-energy individual events, time-integrated studies of the IceCube data suggest statistically significant anisotropies in the spatial distribution of IceCube sky-maps events (see, e.g.\citealt{aartsen_timeintegrated_2020}, \origin, \originII, \citealt{buson_erratum_2022}).
Among others,  time-integrated searches performed on IceCube sky maps (\origin, \originII, see also \citealt{buson_erratum_2022, bellenghi_correlating_2023}), found evidence for statistically significant correlations with the 5th edition of the Roma Blazar catalog \citep[5BZCat;][]{massaro_5th_2015}. Those spatial anisotropies are referred to as neutrino hotspots. The study of the southern celestial hemisphere with 7-yrs sky maps (\origin) and the northern celestial hemisphere with 9-yrs sky maps (\originII) highlighted 10 and 42 spatially correlated blazars with individual neutrino hotspots, respectively. 

\begin{figure}
    \centering
    \includegraphics[width=1\linewidth]{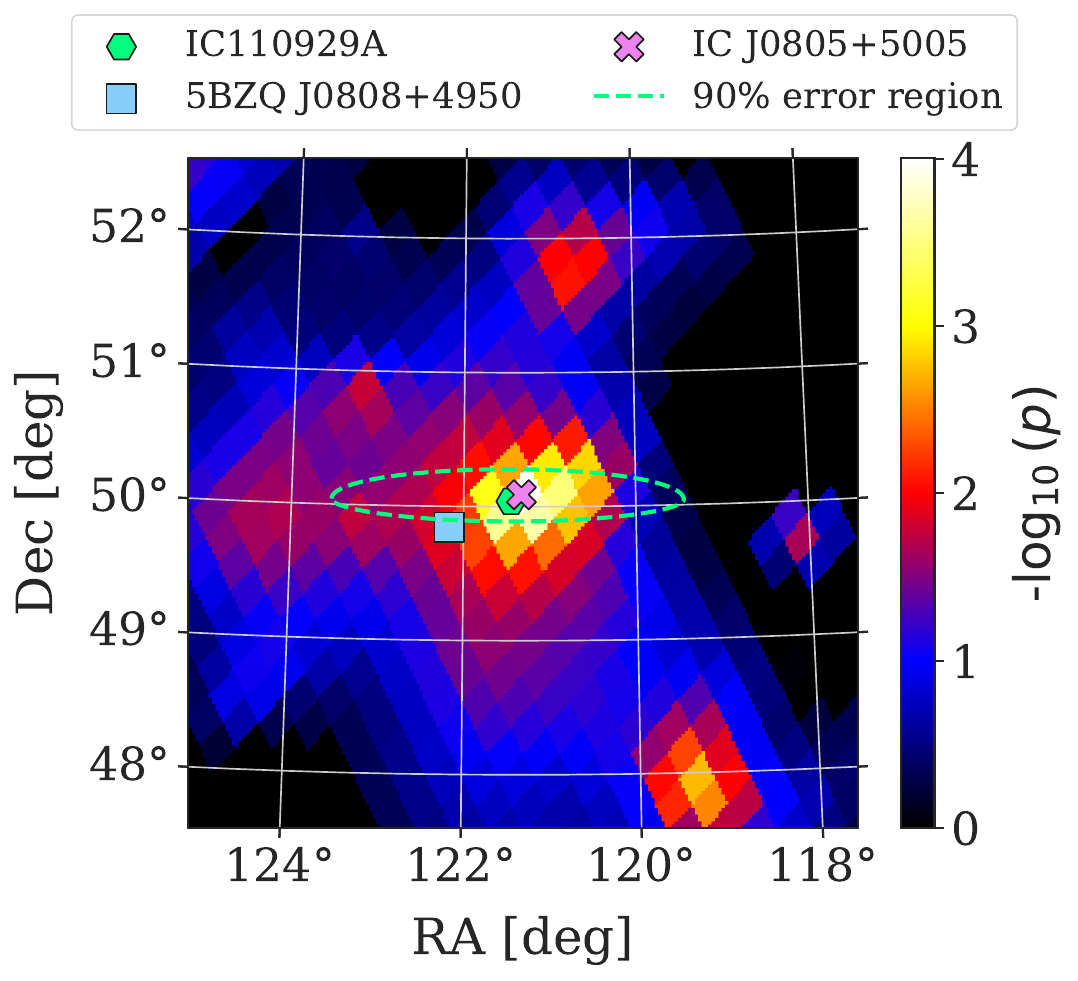}
    \caption{Sky map of the region around the neutrino hotspot reported in \originII. We use the IceCube collaboration data released in \cite{icecubecollaboration_evidence_2022a}. The color map indicates the significance of the neutrino signal (p-value). The position of \target\ is indicated with the cyan square, while the location of the high-energy event reported in IceCat-1 \citep{abbasi_icecat1_2023} is shown with the green hexagone, and the map is centered at the position of the neutrino hotspot indicated by the pink cross. The 90\% error region reported in \cite{abbasi_icecat1_2023} is shown with the green dashed line.}
    \label{fig:neutrino_skymap}
\end{figure}

Among the 52 sources positionally consistent with the neutrino hotspots \target\ ($\mathrm{RA} = 122.16521^\circ,\, \mathrm{Dec} = 49.84361^\circ$), located at a redshift $z = 1.4344$ \citep{shen_catalog_2011}, is tentatively associated with the neutrino hotspot IC J0805+5005 (\originII).  An excess of neutrino events in the direction of \target\, although not significant, was also reported by \cite{smith_revisiting_2021} using three years of IceCube data. Additionally, a high-energy neutrino event \icevent\ from 2011-09-29 is reported in the first IceCube event catalog of alert tracks \cite[IceCat-1;][]{abbasi_icecat1_2023}. \target\ is located right outside the border of the error region derived for this neutrino event. 
Fig. \ref{fig:neutrino_skymap} displays the IceCube neutrino sky map \citep[obtained with the data presented and released in][]{IceCubecollaboration_evidence_2022} centered at the position of the hotspot IC J0805+5005. The position of the blazar \target\ and the high-energy event \icevent\ are shown with the cyan and green crosses, respectively. The pink cross is centered at the position of the hotspot IC J0805+5005 (\originII), and we indicate the significance of the neutrino signal from this direction of the sky with the negative logarithm of the p-value.

\target\ is classified as a flat-spectrum radio quasar \citep[FSRQ,][]{owen_observations_1978}, traditionally differentiated from BL Lacs objects based on the width and intensity of emission lines in their optical spectra \citep{urry_unified_1995, urry_agn_2004}. The emission lines, expected to be strong and broader in the FSRQ spectra, are tracing the accretion mechanism, hinting for efficient accretion, and reprocessing of part of the disk photons by the broad-line region (BLR), and also by the dust torus surrounding the AGN. These reprocessed photons provide additional targets for the interactions with the blazar jet, which can be favorable in order to produce neutrinos \citep{mannheim_neutrinos_1992, mannheim_proton_1993, dermer_photopion_2014, bottcher_leptonic_2013, keivani_multimessenger_2018, petropoulou_multiepoch_2020,zhang_neutral_2020}.

In this work, we collect multi-wavelength data of \target\ and define six time intervals, as detailed in Section \ref{sec:data}. In order to assess the properties of the source over time, and probe whether \target\ can be an efficient neutrino emitter, as hinted by the sky map analysis, we model the spectral energy distribution (SED) of \target\ for each epoch. The jet model and the emission mechanisms considered for this work are described in Section \ref{sec:model}. We present our results in Section \ref{sec:results}, and compare our results with previous works, commenting on the neutrino production efficiency of the source in Section \ref{sec:discussion}. Finally, the main conclusions and summary of this work are presented in Section \ref{sec:conclusion}.

Throughout the paper, we assume a flat $\Lambda\mathrm{CDM}$ cosmology \citep{planckcollaboration_planck_2020}.

\section{MULTIWAVELENGTH OBSERVATIONS AND DATA ACQUISITION}
\label{sec:data}

\begin{figure*}
    \centering
    \includegraphics[width=0.95\linewidth]{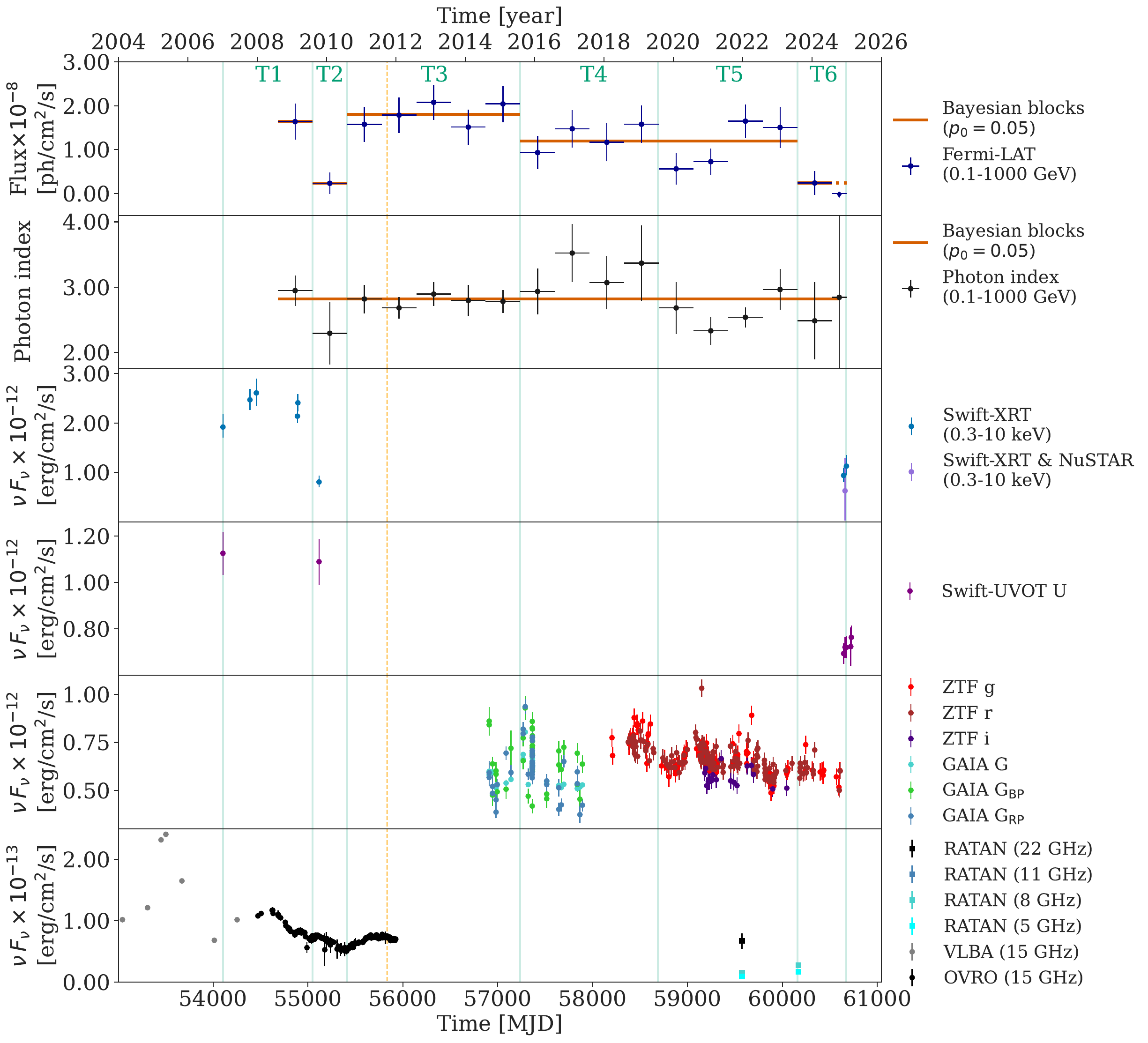}
    \caption{Multi-wavelength light curve of \target\ from $\gamma$-rays to the radio regime, from top to bottom. The green lines separate the time intervals used for the SED modeling. The orange dashed line indicates the arrival time of the high-energy neutrino \icevent. 
    \textbf{Top panel}: $\gamma$-ray light curve with 12-month binning as observed by the \textit{Fermi-}LAT.
    The red horizontal line shows the results of the Bayesian block analysis to identify variable states. 
    \textbf{Second panel}: Same as the first panel, but for the gamma-ray photon index. 
    \textbf{Third panel}: X-rays measurements from \textit{Swift}-XRT and \textit{NuSTAR}, between $0.3 \, \mathrm{keV}$ and $10\,\mathrm{keV}$. 
    \textbf{Fourth panel}: Data from \textit{Swift}-UVOT. 
    \textbf{Fifth panel}: Optical observations from several instruments, corrected for galactic extinction. 
   \textbf{Sixth panel}: Radio measurements with RATAN, OVRO, and VLBA.}
    \label{fig:J0808_LC}
\end{figure*}

\subsection{Fermi-LAT analysis}

Over sixteen years of data provided by the \textit{Fermi}-Large Area Telescope (LAT) are reduced and analyzed with the python package fermipy (v1.2)\footnote{More details at \url{https://fermipy.readthedocs.io/en/latest/}} \citep{wood_fermipy_2017} and the Fermitools (v2.2.0). \target~ is associated with the $\gamma$-ray source \targetgamma\ in the \textit{Fermi}-LAT fourth source catalog \citep[4FGL-DR4;][]{abdollahi_incremental_2022, ballet_fermi_2023}, for which we extract photons in a time interval between 2008-08-04 and 2024-12-31 (MJD 54682-60675) and an energy range of $100\,\mathrm{MeV}$ to $1\,\mathrm{TeV}$ for a region of interest (ROI) of $10^\circ\times10^\circ$ centered on the source. A detailed description on the data handling, the analysis and the fitting procedure can be found in Appendix \ref{Appendix: Fermi}. For the production of the light curve, we kept the flux normalization and the photon index as free parameters and used an integration time of one year per time bin.
The gamma-ray light curve is displayed in the first panel of Fig. \ref{fig:J0808_LC}. The variation of the photon index over time is presented in the second panel of the same figure. Using Bayesian blocks \citep{scargle_studies_2013,theastropycollaboration_astropy_2013} with a false positive rate $p_0=0.05$, we determine changes in flux and photon index. In the flux light curve produced for \targetgamma, we identify five distinct states, while the photon index does not show significant variations. This is consistent with a variability index of $\mathrm{TS}_{\mathrm{var}}=46.01$ (16 degrees of freedom) obtained with fermipy, corresponding to a deviation from a steady state hypothesis at a $\sim3.9\sigma$ level.
For the longest state, we separated the interval into two smaller intervals due to the large errors reported in the spectral index variations, even though they are not significant enough to be reported as a change in the Bayesian blocks analysis.
For each of these six epochs, labeled T1 to T6 (see Fig. \ref{fig:J0808_LC}), we produce a gamma-ray spectrum. 

In epoch T3, we find a significant high-energy spectral point at $\sim 100\,\mathrm{GeV}$. This detection seems to be consistent with a similar trend observed in the SED published in 4FGL-DR4 for \targetgamma. Given that the associated neutrino event \icevent\ happens during the same time interval and that the source displays its highest $\gamma$-ray emission during this epoch, we performed a deeper investigation using the \texttt{gtsrcprob}\footnote{See \url{https://fermi.gsfc.nasa.gov/ssc/data/p7rep/analysis/scitools/help/gtsrcprob.txt} for further details.} method from the Fermi Science Tools. The results reveal that this excess is produced by one HE photon of $\sim129\,\mathrm{GeV}$ detected on 2012-05-12 and is likely associated with the galactic or isotropic diffuse flux (with a probability of $41\%$ and $58\%$, respectively). The probability that this photon is coming from \targetgamma\ is $\sim1\%$. At this energy and this redshift, the optical depth of the extragalactic background light (EBL) is expected to be $\tau_{\gamma\gamma}(E=129\,\mathrm{GeV}, z=1.4344)\approx 2.05-2.99$ for the models of \cite{franceschini_extragalactic_2008,dominguez_extragalactic_2011,finke_modeling_2010}, so a photon signal at this energy would be attenuated by $82\%$ to $95\%$, supporting the fact that this detection is not associated with the blazar.%

\subsection{X-ray observations} \label{section: X-ray}

\begin{table*}
    \begin{adjustbox}{width=\textwidth}
    \begin{tabular}{|c|c|c|c|c|c|c|c|c|c|c|}
         \hline
         Epoch&Obs. Date&Instrument&Obs. ID&Exposure&Rate&$\Gamma$&$N_\mathrm{H,ISM}$&$F_{\mathrm{abs}}$&$F_{\mathrm{int}}$&Cstat/dof\\
           &&&&& $\times10^{-2}$&&$\times10^{21}$&$\times10^{-13}$&$\times10^{-13}$&\\
            &MJD&&&ks& $\mathrm{cts/s}$&&$\mathrm{cm}^{-2}$&$\mathrm{erg}\,\mathrm{cm}^{-2}\,\mathrm{s}^{-1}$&$\mathrm{erg}\,\mathrm{cm}^{-2}\,\mathrm{s}^{-1}$&\\
         \hline
         \multirow{5}{*}{T1} & 54104 & \multirow{5}{*}{XRT} & 00036196001 & 4.1 & $2.9\pm0.3$ & \multirow{5}{*}{$1.7\pm0.1$}& \multirow{5}{*}{$2.6\pm0.9$}& \multirow{5}{*}{$19\pm1.0$}& \multirow{5}{*}{$22\pm1.0$}&\multirow{5}{*}{$368.69/402$}\\
                             & 54389 &                      & 00036196002 & 4.6 & $4.3\pm0.3$ & & & & &\\
                             & 54456 &                      & 00036196003 & 3.9 & $4.3\pm0.3$ & & & & &\\
                             & 54889 &                      & 00036196004 & 9.4 & $3.7\pm0.2$ & & & & &\\
                             & 54894 &                      & 00036196005 & 6.4 & $4.3\pm0.3$ & & & & &\\
         \hline
          T2                 & 55117 & XRT                  & 00036196006 & 5.1 & $1.4\pm0.2$ & $1.8\pm0.2$& $<2.9$ & $7.3\pm1.3$&$8.1\pm1.0$&$64.10/62$\\
          \hline
         \multirow{4}{*}{T6} & 60645 & XRT                  & 00036196007 & 4.8 & $1.4\pm0.2$ & \multirow{4}{*}{$1.5 \pm 0.1$}& \multirow{4}{*}{$<3.7$  }& \multirow{4}{*}{$6.5 \pm 0.3$}& \multirow{4}{*}{$7.7 \pm 0.4$ }&\multirow{4}{*}{$599.30/753$}\\
          &   \multirow{2}{*}{60660} & FPM                  & 61060002002 & 38.9 & $2.1\pm0.1$ & & & & &\\
          &                          & XRT                  & 00097511002 & 4.4 & $1.3\pm0.2$ & & & & &\\
          &                    60674 & XRT                  & 00036196008 & 3.9 & $1.5\pm0.2$ & & & & &\\
          \hline
    \end{tabular}
    \end{adjustbox}
    \caption{X-rays observations of the source \target, and fit results for the \textit{Swift}-XRT and \textit{NuSTAR} observations of \target, including both absorbed ($F_{\mathrm{abs}}$) and intrinsic ($F_{\mathrm{int}}$) fluxes. Uncertainties are reported at the 90\% confidence level within the 0.3–10.0\,keV energy range. In T6, the multiplicative constants accounting for cross-calibration between different instruments are $\mathrm{C}_{\mathrm{XRT,FPMA}}=0.9 \pm 0.1$ and $\mathrm{C}_{\mathrm{XRT,FPMB}}=0.8 \pm 0.1$ respectively.}
    \label{tab:xrays_obs}
\end{table*}

X-ray observations are available from the X-Ray Telescope (XRT) instrument on board the \textit{Neil Gehrels Swift Observatory} (\textit{Swift}-XRT) and the \textit{Nuclear Spectroscopic Telescope Array} (\textit{NuSTAR}). For this work, we make use of both archival data and new observations obtained by our team during \textit{Swift} Cycle 20 (e.g., Proposal Number 2023182), as well as \textit{Swift} Targets of Opportunity (e.g., ToO ID: 20481).

Following \cite{sanchezzaballa_xray_2025}, the \textit{Swift}-XRT data \citep{gehrels_theswiftgammaray_2004} were processed using \textsc{FTOOLS} within the \textsc{HEASoft} package (v6.33)\footnote{https://heasarc.gsfc.nasa.gov/docs/software/lheasoft/}. The \textit{Swift}-XRT observations were performed in photon counting (PC) mode. Event files were calibrated and cleaned using the \texttt{xrtpipeline} task (v.0.13.7), applying standard filtering criteria and the latest calibration files from the \textit{Swift} \textsc{CALDB} maintained by HEASARC. For spectral extraction, we selected events in the $0.3-10.0$~keV energy range using \texttt{xselect}. After visually inspecting the images, we selected the extraction region for the source from a circular region with a radius of 20 pixels ($\sim$47"), which encloses approximately $90\%$ of the XRT point spread function, centered on the source. The background spectrum was obtained from an annular region centered on the source, with an inner radius of 40 pixels ($\sim$1.5') and an outer radius of 150 pixels ($\sim$6.0'). Exposure maps and ancillary response files (ARFs) were generated using the \texttt{xrtexpomap} and \texttt{xrtmkarf} tasks, respectively. Particularly for epochs T1 and T6 that we have multiple XRT pointings, we combined the corresponding event files and exposure maps into single datasets using the \texttt{xselect} and \texttt{ximage} tools for each epoch.

The \textit{NuSTAR} observations, conducted in SCIENCE mode, were processed using the \textit{NuSTAR} Data Analysis Software (\textsc{NuSTARDAS}) v2.1.2 \citep{harrison_nuclear_2013}. Data from both Focal Plane Modules (FPM), A and B, were used in this analysis. The raw event files were calibrated with the \texttt{nupipeline} task, applying the latest calibration files from CALDB v20250602. Following a visual inspection of the event files, the source spectrum was extracted from a circular region centered on the source, with a radius of approximately 50". The background spectrum was derived from an annular region centered on the same position, with inner and outer radii of $\sim$2.0' and $\sim$3.3', respectively. Using the \texttt{nuproducts} script, we then generated the source and background spectra, along with the associated ancillary response files (ARFs) and redistribution matrix files (RMFs).

To enable proper spectral modeling, we applied binning to ensure a minimum of one count per spectral channel, allowing the use of C-statistics \citep{cash_parameter_1979}. Binning was performed using \texttt{grppha} for the \textit{Swift} and \textit{NuSTAR} datasets. This approach is well-suited for low-count data, as the C-statistic is derived from the likelihood function appropriate for Poisson-distributed events. Spectral fitting was carried out using \textsc{XSPEC} (v12.14), with uncertainties estimated at the 1$\sigma$ confidence level \citep{arnaud_xspec_1996}. The Galactic hydrogen column density in the direction of \target\ was fixed at $N_{\text{H, gal}} = 4.67 \times 10^{20}$ cm$^{-2}$ \citep{hi4picollaboration:_hi4pi_2016}
. The metal abundances were set to Solar values, adopting those from \cite{wilms_absorption_2000}
, and the photoelectric cross-sections for all absorption components were taken from \cite{verner_atomic_1996}. 
Additionally, we included, for all tested models, an absorption component at the redshift of the source ($N_{\text{H, ISM}}$, fixing $z = 1.4344$) to account for interstellar medium (ISM) absorption \citep{abazajian_seventh_2009}. For the T6 epoch, which includes a simultaneous observation, we included a multiplicative constant in the spectral model to account for cross-calibration uncertainties between different instruments \citep{madsen_calibration_2015}. Specifically, the \textit{NuSTAR} FPMA and FPMB spectra were analyzed in the 3.0–30.0 keV energy range, as above this energy range, the spectra were background-dominated. In the case of \textit{Swift} observation, we included and fitted the data in the 0.3–10 keV range. 

We modeled the spectra using the simplest spectral shape: a power-law with photon index $\Gamma$, including Galactic absorption and an additional absorption component from the ISM, following the model \texttt{phabs*zphabs*pow}. For the two tested jet models, we combine all data that correspond to each state. The results of the fit are presented in Table \ref{tab:xrays_obs}. Alternative models, such as a broken power-law or a log-parabola, were also tested, but they did not provide an improvement to the fit or evidence for an additional spectral component at the 1$\sigma$ significance level. Each observation was analyzed and fitted individually. To construct the average SED model and plot it, all individually fitted datasets were rebinned to a significant detection as large as $5\sigma$. Additionally, for the SED modeling, a 20\% fractional systematic uncertainty was added to the data above $\sim$10 keV, since information in this energy range is only available during the T6 state, when \textit{NuSTAR} performed a single observation. This takes into account, in a conservative way, possible short and long-term hard X-ray variability, that can reach values well above $20\%$ \citep[see, e.g.,][]{costamante_nustar_2018, mundo_longterm_2023} in blazars.

\subsection{Infrared, optical and UV observations}
\label{sec:optical spectro}

Spectroscopic observations in the optical are also available for \target\ in the Seventh Data Release of the Sloan Digital Sky Survey \citep[SDSS-DR$7$, analyzed in][]{shen_catalog_2011}. These provided a measure of the redshift based on the detection of a broad \MgII\ emission line consistent with the value $z = 1.4344$. We used this spectral profile and adopted the relation between the luminosities of the emission lines and the BLR based on the composite blazar spectrum of \cite{francis_high_1991} to infer the $L_{\rm BLR}$ of the source \citep[see a similar approach in][hereafter \paperPHone]{sbarrato_relation_2012, azzollini_physical_2025}. We assumed a covering factor of $0.1$ for the BLR relative to a geometrically thin accretion disk and estimated the disk luminosity. Then, we estimated the virial mass of the central black hole adopting the empirical scaling laws of \cite{shen_catalog_2011, shen_comparing_2012}. The resulting values are  $L_{\rm disk} = \left(1.35 \pm 0.25\right)\times10^{46}\,{\rm erg}\,{\rm s}^{-1}$ and $M_{\rm BH}=\left(6.95\pm0.13\right)\times10^{8}\,M_{\odot}$ (\paperPHone; see also \cite{shen_catalog_2011, paliya_central_2021}). 

There are photometric observations available for \target~by the Zwicky Transient Facility \citep[ZTF;][]{masci_zwicky_2018}, Gaia \citep{gaiacollaboration_gaia_2016, gaiacollaboration_gaia_2023,babusiaux_gaia_2023} and the Ultra-Violet Optical Telescope on board 
\textit{Swift} \citep[\textit{Swift-}UVOT;][]{roming_swift_2005}. While additional surveys also include this source in their catalogs, most are significantly contaminated by the nearby star 2MASS J08084034+4950387, located within $<7\,\mathrm{arcsec}$. \textit{Swift-}UVOT observations are available only in the U filter simultaneous to X-ray observations discussed in \ref{section: X-ray}. Given the redshift of the source, Lyman-alpha absorption is expected to suppress flux below  $\sim 2960\,$\AA, affecting all three UVOT UV filters. The source counts are extracted with a radius of $3-5"$, while the background counts are taken from a region around the source position, avoiding visible contamination by the nearby star or other sources in the vicinity. The reddened magnitudes are then computed with the UVOTSOURCE tool provided by HEASOFT \footnote{For more details see \url{https://heasarc.gsfc.nasa.gov/docs/software/heasoft/}}. Galactic extinction corrections for this and subsequent optical instruments are applied following \cite{schlafly_measuring_2011}, using the source-dependent $E
(B-V)$ and assuming a standard reddening law with 
$R_V=3.1$. Wavelength-dependent corrections follow \cite{fitzpatrick_correcting_1999}. The magnitudes are finally converted to energy flux units with the zero-points taken from \cite{poole_photometric_2008} and \cite{breeveld_updated_2011}. Gaia photometry is available in the G, $\mathrm{G}_{\mathrm{BP}}$ and $\mathrm{G}_{\mathrm{RP}}$ bands. After applying the same reddening correction procedure, these are converted to energy flux units using the zero points and the filter information from \cite{riello_gaia_2021}. Lastly, photometric observations performed with ZTF are available in the red (r), the green (g), and the near infrared (i) band with the filter information provided by the Spanish Virtual Observatory Filter Profile Service \citep[SVO;][]{rodrigo_svo_2012,rodrigo_svo_2020}. The light curves in energy flux units and uniformly corrected for galactic extinction are displayed in the third and fourth panels of \ref{fig:J0808_LC} and used for the modeling when they are contemporaneous to the time intervals T1-T6.

\subsection{Radio observations}
\label{sec:radio}

Radio light curves are available from the Academy of Sciences Radio Telescope (RATAN)\footnote{For more details see \url{https://www.sao.ru/blcat/}}, the Very Long Baseline Array \citep[VLBA,][]{lister_mojave_2009} and the Owens Valley 
Radio Observatory 40 m telescope \citep[OVRO,][]{richards_blazars_2011} as published in \cite{richards_connecting_2014}. They are displayed in the bottom panel of Fig. \ref{fig:J0808_LC}. Data at a frequency lower than $10^{11}\,\mathrm{Hz}$ are used as upper limits in the modeling, since this emission likely corresponds to extended regions further away from the core, which we do not consider in our model \citep[see, e.g.][]{konigl_relativistic_1981, ghisellini_general_2010, tavecchio_constraints_1998}. 

\section{SED MODELING}
\label{sec:model}

The SED of blazars is traditionally composed of two distinct components \citep{fossati_unifying_1998}. The low-energy component is commonly attributed to synchrotron radiation from relativistic electrons present in the jet, while the origin of the second hump remains debated. In purely leptonic models, the high-energy part of the SED is attributed to the inverse-Compton scattering (IC) of photons by the electrons of the jet. When considering lepto-hadronic models, in the context of (ultra) high-energy cosmic rays and/or neutrino production, the high-energy component can be explained in different ways: synchrotron radiation from the relativistic protons of the jet (synchrotron proton blazar), or a leptonic-dominated origin and a cascade component \citep{mannheim_proton_1993,aharonian_tev_2000,mucke_proton_2001,cerruti_hadronic_2015, petropoulou_photohadronic_2015}.

\subsection{Jet models}


We consider the case in which the emission region (hereafter "blob") is traveling over time, within a cylindrical jet (i.e., the blob is not expanding over time). We model the blob as a spherical, magnetized region of radius $R'_\mathrm{blob}$. 
We define the emission region's bulk velocity $\beta_\mathrm{j} = \sqrt{1 - \Gamma_\mathrm{j}^{-2}}$, and the jet makes an angle $\theta \sim 1/\Gamma_\mathrm{j}$ (typical for blazars) with the line-of-sight, defining a Doppler factor $\delta_\mathrm{j} = [\Gamma_\mathrm{j}(1 - \beta_\mathrm{j}\cos{\theta_\mathrm{j}})]^{-1}$. With the small line-of-sight angle, we use: $\delta_\mathrm{j}= \Gamma_\mathrm{j}.$ 

Primary particles, namely relativistic electrons and protons, are continuously injected into the region with a constant injection rate $Q'_i (E'_i) = dN'/(dE'dt'dV')$ following a power-law with an super-exponential cutoff 
\begin{equation}
    Q'_i (E'_i) \propto E'^{-p_i}_i \exp{\left[-\left(\frac{E'_i}{E'_{i,\mathrm{max}}}\right)^2\right]} \, \, \text{for $E'_i > E'_{i,\mathrm{min}}$}
\end{equation}
where $V' \int E'\,Q'\,dE' = L'_\mathrm{inj}$ gives the normalization of the injection rate expressed as the injection luminosity. The super-exponential cutoff can be related to diffusive shock acceleration in the Bohm limit \citep{zirakashvili_analytical_2007}.

We use the software package\footnote{v1.2.0} AM$^3$ \citep{gao_modelling_2019, klinger_am3_2023, klinger_am3_2024} to compute the lepto-hadronic interactions in the emission region. The processes taken into account are synchrotron radiation, inverse-Compton scattering, Bethe-Heitler pair production, photo-pion production, proton-proton interaction, and photon-photon annihilation. Particle escape is taken into account, with an escape time $t_\mathrm{esc} = R'_\mathrm{blob}/c$, with $c$ the velocity of light. Details about the interactions, the computational methods, and the software can be found in \cite{klinger_am3_2024}.

For each quasi-simultaneous observation, we evolve the simulation over the duration of the observational periods indicated in the light curve  (Fig. \ref{fig:J0808_LC}, namely an observed time of $1$, $1$, $4.9$, $4$, $4$, and $1.25$ years respectively), corresponding to a time $t' = \delta_\mathrm{j}\,t/(1+z)$ in the rest frame of the jet. 
The blob is injected at a distance $R_\mathrm{diss}$ from the black hole, and the emission region travels along the jet, to distances of the order of $R_\mathrm{diss} + \beta_\mathrm{j}c \Gamma_\mathrm{j}t'$ in the stationary frame of the AGN. 
We calculate the time-averaged SED, and transform it into the observer's frame as $F_\nu = (1+z) \, \delta_\mathrm{j}^3 \, L'_\nu / (4 \pi \, d_L^2)$, where $d_L$ is the luminosity distance, and the frequencies are related through $\nu = \delta_\mathrm{j} \nu' / (1+z)$.

\subsection{External photons fields}

\target\ belongs to the category of the high-excitation radio galaxies (HERGs, \paperPHone), so the accretion mechanisms are expected to be important \citep[see, e.g][]{ghisellini_transition_2011,sbarrato_relation_2012, best_fundamental_2012}. 
We take into account photons from the accretion disk, the broad-line region, and we also consider the presence of a dusty torus in the vicinity of the supermassive black hole. The transformation of these external photon fields into the jet frame in order to carry on the radiative calculations follows \cite{ghisellini_origin_1996} and \cite{ghisellini_canonical_2009}. Depending on the position of the emission region relative to the photon fields, parametrized as $r_\mathrm{diss} = R_\mathrm{diss}/R_\mathrm{BLR}$, the photons will be seen as boosted or de-boosted in the jet frame. 
Empirical relations are used to derive the characteristics of the external field in order to reduce the number of free parameters in the models. Specifically, we use: 
\begin{align*}
L_\mathrm{disk} &= 0.1 \, \dot M c^2,\\
L_\mathrm{BLR} &= 0.1 \, L_\mathrm{disk}\,\, \text{\citep[see e.g,][]{ghisellini_origin_1996},}\\
L_\mathrm{torus} &= f_\mathrm{torus}\,L_\mathrm{disk}, \\
\end{align*}
where $\dot{M}$ is the mass accretion rate of the disk, $L_\mathrm{disk}$, $L_\mathrm{BLR}$ and $L_\mathrm{torus}$ are the total disk, BLR and torus luminosities respectively and $f_\mathrm{torus}$ is the covering factor of the torus, that we fix to $f_\mathrm{torus} = 0.5$ \citep{calderone_widefield_2012, gu_evolution_2013}.
Furthermore,
\begin{eqnarray}
R_\mathrm{BLR} &=& 10^{17} \, \left(\frac{L_\mathrm{disk}}{10^{45} \, \mathrm{erg/s}}\right)^{1/2} \, \mathrm{cm}\\
R_\mathrm{torus} &=& 1.2\times10^{18} \, \left(\frac{L_\mathrm{disk}}{10^{45} \, \mathrm{erg/s}}\right)^{1/2} \, \left(\frac{T_\mathrm{torus}}{10^3 \, \mathrm{K}}\right)^{-2}  \, \mathrm{cm}.
\label{eq:torus}
\end{eqnarray}
We let the torus temperature be a free parameter, but enforce that it always stays below the dust sublimation temperature $T_\mathrm{torus} \lesssim 1300\mathrm{K}$.

 Effectively, the accretion disk is modeled as a multi-temperature blackbody, where we assume that the disk is perpendicular to the jet axis \citep{frank_accretion_2002}, and extending between the inner radius $R_\mathrm{in} = r_\mathrm{in} \, G M_\mathrm{BH}/c^2 = r_\mathrm{in} \, R_\mathrm{g}$, with $G$ the gravitational constant, where $R_\mathrm{g}$ is the gravitational radius of the black hole and the outer radius $R_\mathrm{out} = 500 \,R_\mathrm{in}$: 
 \begin{equation*}
     F_\nu^\mathrm{AD}(\nu) =   \frac{4\pi h \nu^3}{c^2 d_L^2} \int_{R_\mathrm{in}}^{R_\mathrm{out}} \frac{R dR}{\exp{[h\nu/k_B T(R)] - 1}},
 \end{equation*}
where the temperature $T(R)$ is given by the standard relation for thin disks \citep[e.g.][]{frank_accretion_2002}
\begin{equation*}
T(R) = \left[\frac{3 G \dot{M} M_\mathrm{BH}}{8 \pi R^3 \sigma}\left(1 - \sqrt{\frac{R_\mathrm{in}}{R}}\right)\right]^{1/4},
\end{equation*}
$h$ is the Plank constant, $k_B$ is the Boltzmann constant and $\sigma$ is the Stefan-Boltzmann constant. 

Additionally, the spectrum of the torus is described as a single-temperature blackbody, normalized to the luminosity of the torus 
\begin{equation*}
    \nu\, F_\nu^\mathrm{DT}(\nu) = \frac{15 L_\mathrm{torus}}{4\pi^5 d_L^2} \left(\frac{h\nu}{k_B T_\mathrm{torus}}\right)^4 \left[\exp{\left(\frac{h\nu}{k_B T_\mathrm{torus}}\right) - 1}\right]^{-1}
\end{equation*}

\section{Results}
\label{sec:results}

We model the quasi-simultaneous epochs with the procedure described above. As described in Section \ref{sec:radio}, the SED contains archival radio data, which we treat as upper limits, as the low-frequency emission from the jet is likely originating from the extended jet. We fix the black hole mass and the redshift of the source to the values derived in the literature (\paperPHone). 
We scan parameters with a least-square minimization algorithm, and assess the fit quality with the chi-square value defined as 
\begin{equation}
\label{eq:chi2}
\chi^2_\mathrm{red} = \frac{1}{\mathrm{d.o.f}} \sum_{i=1}^{N}\frac{(y_i - f(x_i))^2}{\sigma_i^2}
\end{equation}
where $N$ is the number of data points, $(x_i, y_i, \sigma_i=\sigma_{y,i})$ represent the data points and the uncertainty on the $y_i$ value (i.e., the error bar), respectively, and $f(x_i)$ is the model prediction. The number of degrees of freedom, d.o.f, is determined by the number of data points minus the number of free parameters in the models. Since the frequency bins span very different scales depending on the frequency, we also compute an effective chi-square, where we replace variance $\sigma_i$ by an effective one such that $ \sigma^2_i \rightarrow \sigma^2_{i, \mathrm{eff}} = \left[ \frac{\partial f}{\partial x}\right]_i^2\sigma_{x,i}^2 + \sigma_{y,i}^2$ \citep{orear_least_1982, orear_erratum_1984}, in order to take into account the uncertainty in the simulated flux induced by the finite bin width. The upper limits in the data are not used in the calculation of the degrees of freedom, but we consider them in the fit so that the predicted flux never violates them (a weigth of $10^5$ is added to the residual of a certain point if it violates an upper limit). We start the fitting procedure with the epochs T1, T2, and T6, where the X-ray observations are available, and lead to a number of degrees of freedom $\mathrm{d.o.f} \ge 1$. For T3, T4, and T5, the number of free parameters is larger than the number of fitted data points. To limit this degeneracy we use the solutions obtained in T1, T2, and T6 as starting points for the fitting procedure of the remaining epochs. However, the goodness of fit in the periods T3, T4, and T5 cannot be strictly compared to the epochs T1, T2, and T6, due to the large number of parameters compared to the number of data points (we report the $\chi^2$ value, the number of data points and if applicable, the number of degrees of freedom in the last row of Table \ref{tab:state_parameters}).

In order to reduce the number of free parameters we make standard assumptions, following, e.g, \cite{cerruti_hadronic_2015}, the minimum energy of the injected protons is set to $\gamma_p' = 1.0$, and the proton index to $p_\mathrm{p} = p_e$. With this, we assume that the electrons and the protons are accelerated via the same mechanism while they are injected into the emission region. Setting $\gamma_p' = 1$ is also a conservative choice in terms of the resulting jet power because a larger value would lead to a lower proton density needed to obtain equal results. 

We let the accretion rate $\dot{M}$ vary within the uncertainty of the value inferred in \paperPHone. The ranges of parameters that we explore is presented in the last column of Table \ref{tab:state_parameters}, along with the results of the modeling.
We report the best-fit parameters of the simulations, as well as some derived quantities such as the accretion regime $\eta=L_\mathrm{BLR}/L_\mathrm{Edd}$ or the jet power $P_\mathrm{jet} = 2\pi\,\Gamma_\mathrm{j}^2\beta_\mathrm{j}R'^{2}_\mathrm{blob}c\sum u'_i$, where we sum the energy density $u'_i$ of the magnetic field $u'_\mathrm{b}=B'^2/(8\pi)$, the electrons $u'_\mathrm{e}=L'_\mathrm{e}/(4\pi R'^2_\mathrm{blob}c)$, the protons $u'_\mathrm{p}=L'_\mathrm{p}/(4\pi R'^2_\mathrm{blob}c)$, and the radiation $u'_\mathrm{rad}$.
We explore a broad range of parameters, either limited by physical constraints (such as the torus temperature or the accretion rate), or in order to scan different types of solutions (e.g., compact blobs with sizes $R'_\mathrm{blob} < 10^{16}\,\mathrm{cm}$, or larger magnetic fields strengths $B' > 15\,\mathrm{G}$. With the large number of free parameters (13, see Table \ref{tab:state_parameters}), other solutions may also fit the SEDs, but we report here the best ones we found with a non-negligible neutrino output and a consistent evolution throughout the six epochs.

\begin{table*}
\begin{adjustbox}{width=\textwidth}
\renewcommand{\arraystretch}{1.3}
\begin{tabular}{|l|c|c|c|c|c|c|c|c|} 
\hline
$-$ & unit & T1 & T2 & T3* & T4  & T5 & T6 & explored range \\
\hline
Start & $-$ & 2008-08-04 & 2009-08-04 & 2010-08-04 & 2015-08-03 & 2019-07-23 & 2023-08-01 & $-$ \\
End & $-$ & 2009-08-04 & 2010-08-04 & 2015-08-03 & 2019-07-23 & 2023-08-01 & 2024-12-31 & $-$ \\
\hline
$R'_{\rm blob}$ & $\rm{cm}$ & $1.0^{+0.2}_{-0.2} \times 10^{16}$ & $3.0_{-0.1}^{+0.3} \times 10^{16}$ & $3.0 \times 10^{16}$ & $2.0 \times 10^{16}$ & $2.8 \times 10^{16}$ & $1.5_{-0.2}^{+0.09} \times 10^{16}$ & $[5\times10^{15},1\times10^{17}]$\\
$B'$ & $\rm{G}$ & $3.1^{+0.5}_{-0.4}$ & $5.0_{-0.4}^{+2.8}$ & $8.5 \times 10^{-1}$ & $5.4\times 10^{-1}$ & $1.0$ & $2.5_{-0.3}^{+0.1}$ & $[10^{-1},5\times10^{1}]$\\
$\delta_\mathrm{j}$ & $-$ & $5.5^{+0.1}_{-0.1}$ & $5.5_{-0.3}^{+0.07}$ & $5.3$ & $5.3$ & $5.3$ & $5.1_{-0.07}^{+0.03}$ & $[5.0,3\times10^{1}]$\\
$\gamma'_{\rm e, min}$ & $-$ & $7.0^{+0.5}_{-0.3} \times 10^{2}$ & $6.6_{-6}^{+23} \times 10^{1}$ & $1.7 \times 10^{2}$ & $7.0 \times 10^{3}$ & $1.2 \times 10^{2}$ & $8.0_{-0.7}^{+0.1} \times 10^{2}$ & $[1.0,10^4]$\\
$\gamma'_{\rm e, max}$ & $-$ & $2.5^{+0.3}_{-0.3} \times 10^{4}$ & $2.3_{-0.6}^{+1.5} \times 10^{4}$ & $1.9 \times 10^{4}$ & $1.0 \times 10^{4}$ & $4.0 \times 10^{4}$ & $1.9_{-0.2}^{+0.2} \times 10^{4}$ & $[10^{3},5\times10^{5}]$\\
$p_{\rm e}$ & $-$ & $1.9^{+0.04}_{-0.05}$ & $2.0_{-0.03}^{+0.05}$ & $2.1$ & $1.9$ & $2.2$ & $2.0_{-0.03}^{+0.09}$ & $[1.5,3.0]$\\
$L'_{\rm e}$ & $\rm{erg}\,\rm{s}^{-1}$ & $1.9^{+0.1}_{-0.1} \times 10^{44}$ & $6.1_{-0.6}^{+0.2} \times 10^{43}$ & $2.9 \times 10^{44}$ & $1.1 \times 10^{44}$ & $2.5 \times 10^{44}$ & $8.0_{-0.4}^{+0.3} \times 10^{43}$ & $[10^{40},10^{45}]$\\
$\gamma'_{\rm p, min}$ & $-$ & $1.0$ & $1.0$ & $1.0$ & $1.0$ & $1.0$ & $1.0$ & fixed \\
$\gamma'_{\rm p, max}$ & $-$ & $6.0^{+2}_{-4} \times 10^{6}$ & $1.0_{-0.4}^{+0.2} \times 10^{7}$ & $8.4 \times 10^{6}$ & $4.0 \times 10^{6}$ & $8.3 \times 10^{6}$ & $1.0_{-0.6}^{+0.3} \times 10^{7}$ & $[10^{5},10^{8}]$\\
$p_{\rm p}$ & $-$ &$1.9^{+0.04}_{-0.05}$ & $2.0_{-0.03}^{+0.05}$ & $2.1$ & $1.9$ & $2.2$ & $2.0_{-0.03}^{+0.09}$ & fixed \\
$L'_{\rm p}$ & $\rm{erg}\,\rm{s}^{-1}$ & $1.7^{+0.4}_{-0.9} \times 10^{46}$ & $5.4_{-2.6}^{+1.0} \times 10^{46}$ & $1.3_{-0.5}^{+2.6} \times 10^{47}$ & $1.5_{-1.5}^{+0.0} \times 10^{46}$ & $4.9_{-4.9}^{+9.8} \times 10^{46}$ & $5.5_{-3.3}^{+1.8} \times 10^{46}$ &  $[0,10^{48}]$\\
$\dot{M}$ & $M_\odot\,\rm{yr}^{-1}$ & $2.60^{+0.22}_{-0.66}$ & $2.82_{-0.88}^{+0.0}$ & $2.80$ & $2.75$ & $2.82$ & $2.82_{-2.3}^{+0.0}$ & $[1.94,2.83]$\\
$R_{\rm in}$ & $R_{\rm g}$ & $4.5^{+17}_{-1.5}$ & $5.0_{-2.0}^{+15}$ & $4.7$ & $6.7$ & $4.9$ & $4.5_{-0.7}^{+0.3}$ & $[3.0,5\times 10^1 ]$\\
$R_{\rm diss}/R_{\rm BLR}$ & $-$ & $7.7^{+92.3}_{-5.7} \times 10^{-1}$ & $7.7_{-5.7}^{+92.3} \times 10^{-1}$ & $7.7 \times 10^{-1}$ & $7.7 \times 10^{-1}$ & $7.7 \times 10^{-1}$ & $7.7_{-5.7}^{+92.3} \times 10^{-1}$ & $[5\times 10^{-1},10^1]$\\
$T_{\rm torus}$ & $\rm{K}$ & $1.0^{+0.3}_{-0.9} \times 10^{3}$ & $1.0_{-0.9}^{+0.3} \times 10^{3}$ & $1.0 \times 10^{3}$ & $1.0 \times 10^{3}$ & $1.0 \times 10^{3}$ & $1.0_{-0.9}^{+0.3} \times 10^{3}$ & $[10^{2},1.3\times10^{3}]$\\
\hline
$u'_{\rm e}/u'_{\rm B}$ & $-$ & $1.2 \times 10^{1}$ & $1.8 \times 10^{-1}$ & $3.0 \times 10^{1}$ & $6.7 \times 10^{1}$ & $2.1 \times 10^{1}$ & $3.8$ & $-$ \\ 
$u'_{\rm p}/u'_{\rm B}$ & $-$ & $1.0 \times 10^{3}$ & $1.6 \times 10^{2}$ & $1.3 \times 10^{4}$ & $8.7 \times 10^{3}$ & $4.1 \times 10^{3}$ & $2.6 \times 10^{3}$ & $-$ \\ 
$L_{\rm disk}$ & $\rm{erg}\cdot\rm{s}^{-1}$ & $1.5 \times 10^{46}$ & $1.6 \times 10^{46}$ & $1.6 \times 10^{46}$ & $1.6 \times 10^{46}$ & $1.6 \times 10^{46}$ & $1.6 \times 10^{46}$ & $-$ \\ 
$\eta$ & $-$ & $1.7 \times 10^{-2}$ & $1.8 \times 10^{-2}$ & $1.8 \times 10^{-2}$ & $1.8 \times 10^{-2}$ & $1.8 \times 10^{-2}$ & $1.8 \times 10^{-2}$ & $-$ \\ 
$P_{\rm jet}^{\rm inj}$ & $\rm{erg}\cdot\rm{s}^{-1}$ & $2.5 \times 10^{47}$ & $8.1 \times 10^{47}$ & $1.8 \times 10^{48}$ & $2.1 \times 10^{47}$ & $6.9 \times 10^{47}$ & $7.0 \times 10^{47}$ & $-$ \\ 
$P_{\rm jet}^{\rm inj}/L_{\rm Edd}$ & $-$ & $2.9$ & $9.2$ & $2.1 \times 10^{1}$ & $2.4$ & $7.9$ & $8.0$ & $-$ \\ 
$CD$ & $-$ & $4.2$ & $2.8 \times 10^{-1}$ & $6.9$ & $9.5$ & $6.9$ & $2.1$ & $-$ \\ 
$N_{<\nu_\mu+\bar{\nu}_\mu>}$ & $-$ & $0.008^{+0.002}_{-0.004}$ & $0.003^{+0.001}_{-0.002}$ & $0.101^{+0.034}_{-0.039}$ & $0.001^{+0.000}_{-0.001}$ & $0.053^{+0.111}_{-0.053}$ & $0.006^{+0.003}_{-0.004}$  & $-$ \\
\hline
$\chi^2\,(N_\mathrm{points}^\mathrm{fitted})$ & $-$ & $8.4\,(17)$ & $8.5\,(14)$ & $1.0\,(5)$ & $1.3\,(5)$ & $2.6\,(6)$ & $8.1\,(22)$ & -\\
\hline
\end{tabular}
\end{adjustbox}
\caption{Best-fit parameters of the models for the six states of \target. The last column displays the range of parameters explored during the fit. In total, there are 13 parameters that are allowed to vary, against 5 fixed parameters (namely the minimum proton energy, the proton injection index, the BLR and torus covering factors, and the black hole mass). The second-to-last row gives the number of neutrinos $N_{<\nu_\mu+\bar{\nu}_\mu>}$ per epoch predicted by the best-fit models. The explored allowed range of proton luminosity is also reported in terms of number of neutrinos for each epoch. The * indicates the time interval in which the IceCube high-energy neutrino event was detected. The last row reports the chi-square value for each epoch and the corresponding number of fitted data points (i.e the data points above $10^{11}\,\mathrm{Hz}$ and the points that are not upper limits).}
\label{tab:state_parameters}
\end{table*}

\begin{figure*}
    \centering
    \includegraphics[width=\linewidth]{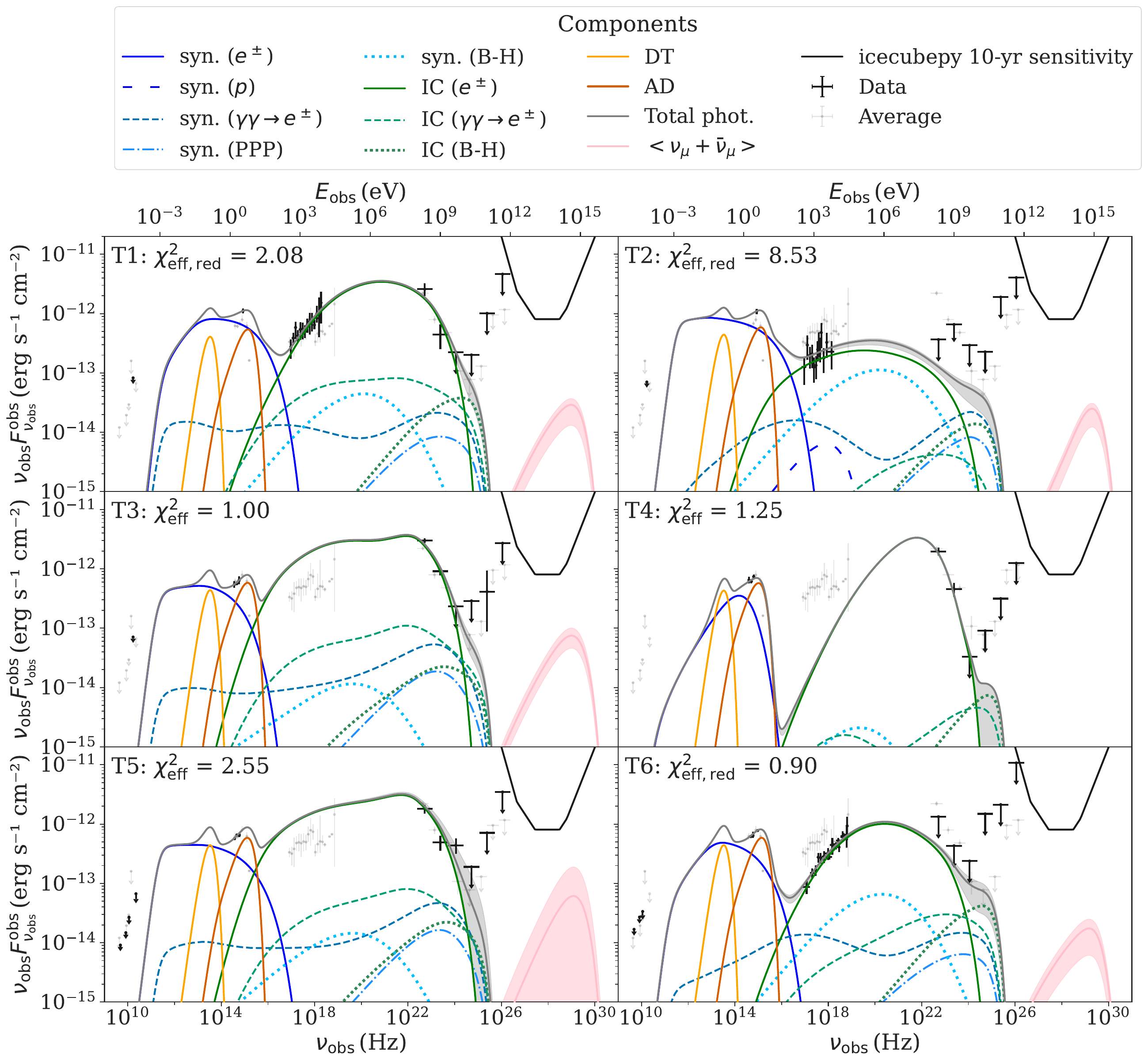}
    \caption{The modeled SEDs of \target, in the observer frame. Each panel corresponds to an epoch, labeled on the top left of the panel. The black points are the contemporaneous points of the studied epoch, while the gray points are the data from the average SED, for reference. Blue lines correspond to synchrotron radiation, while green lines represent inverse-Compton scattering. The processes or particles producing each line are indicated with different line styles, namely solid for primary electrons, loosely dashed for primary protons (visible only in T2, subdominantly around $10^{18}\,\mathrm{Hz}$), dashed for photon-photon pair production, dash-dotted for photo-pion production (PPP), and dotted for Bethe-Heitler pair-production (B-H). In red and orange, we show the emission from the accretion disk and the torus. The total observed photons are shown by the gray line, and the pink line indicates the muon plus anti-muon neutrino flux. The shaded areas correspond to the $1\sigma$ uncertainty on the proton luminosity in the models. We report the \textsf{icecubepy} (\paperNUone, \citealt{Lincetto_in_prep}) 10-year sensitivity at the declination of the source with the dashed black line. The goodness-of-fit is reported as the effective chi-square value reduced to the number of degrees of freedom when applicable}.
    \label{fig:J0808_SED_blob}
\end{figure*}

We show the six SED models, in the observer frame, in Fig. \ref{fig:J0808_SED_blob}. Each panel shows the total time-averaged SED with the solid gray line. The primary and secondary synchrotron radiation components are shown in blue, where the solid line corresponds to primary electrons' synchrotron radiation, the dashed line to the synchrotron radiation from electron-positron pairs created through photon-photon absorption, the dash-dotted one to the synchrotron emission from the photo-pion production secondaries, and the dotted line shows the synchrotron emission from the Bethe-Heitler pair production secondaries. The corresponding inverse-Compton scattering contribution is shown in green, with the same line style for the same process. The red and orange lines indicate the accretion disk and the dusty torus emission, respectively. 
The shaded areas correspond to the range of proton luminosities that fit the data (both SED and light curve points) and do not violate the SED upper limits, or if this is not a constraint, we vary the parameter until $\Delta \chi^2_\mathrm{eff} = 1$. For the three epochs in which we have sufficient data, i.e. T1, T2 and T6, where $\mathrm{d.o.f} \ge1$, we repeat the same procedure as for the proton luminosity for the 13 free parameters independently, and report the corresponding ranges in Table \ref{tab:state_parameters}. We emphasize on the fact that these range are not equivalent to statistical uncertainties, but are shown here to demonstrate which parameters are the most constrained by the data.

The gray data points are the average SED of the source, over the whole time range shown in the light curve in Fig. \ref{fig:J0808_LC}. 
The pink solid line indicates the muon neutrino flux, and we also report the IceCube 10-year sensitivity envelope, obtained with the software \textsf{icecubepy} at the declination of the source with the solid black line (\citet{lincetto_investigating_2025}, hereafter \paperNUone, see also \citealt{Lincetto_in_prep}). 

The SED models are mostly constrained by the UV data points, tracing the interplay between the accretion disk and the jet, and the upper limits in the gamma-ray spectrum, tracing the contribution of the hadronic cascade. The gamma-ray spectral shape, in turn, specifically constrains the neutrino output of a given model. This can be seen particularly in T4, where, despite the large gamma-ray flux, the resulting neutrino output is low. The protons initiate a cascade of interactions, producing secondary particles that in turn emit mainly X-ray photons (through Bethe-Heitler pair production, dotted blue line in Fig. \ref{fig:J0808_SED_blob}), but also above GeV energies (from photon annihilation, photo-pion pair production and Bethe-Heitler pair production). At the redshift of the source, the EBL photons are responsible for the attenuation of the gamma-ray flux above $E_\mathrm{crit} \approx 170\,(1+z)^{-2.38} \approx 20\,\mathrm{GeV}$ \citep{franceschini_extragalactic_2008, ackermann_imprint_2012}. Hence, a non-detection of photons between $1$ and $20\, \mathrm{GeV}$ provides a strong constraint for the models, as the attenuation of the flux must be intrinsic to the source.
The X-ray points, when they are available, constrain the inverse-Compton component, and we are able to explain the X-ray spectral changes between epochs T1, T6 and T2 by a transition from Compton-dominated states (T1 and T6) to a synchrotron-dominated state in T2. When X-ray data are not available, we do not constrain the X-ray flux in any specific way, and we find that our solutions lie within a factor 3 from the flux observed in epoch T1 and below a factor 10 from the lowest flux, observed in T2. These variations are driven only by the models, and are well within long-term variability observed in blazars \citep[e.g. with \textit{Swift}][]{giommi_xray_2021}.

The size of the emission regions is similar for all epochs $R'_\mathrm{blob} \approx (1-3)\times10^{16} \, \mathrm{cm}$. The magnetic field strength varies from $\sim 0.5$ to $5 \, \mathrm{G}$, which is the main variation in the emission region parameters, as the Doppler factor also stays constant around $\delta_\mathrm{j} \approx 5$. \cite{finke_properties_2019} found a magnetic field strength, assuming equipartition between the electrons and the magnetic field, at a distance of $1\,\mathrm{pc}$ from the core of $B'_\mathrm{1pc}=0.9^{+12.0}_{-0.8}\,\mathrm{G}$. Assuming magnetic flux conservation and if the magnetic field (with a toroidal topology) stays in equipartition with the particles, $B'(R'_\mathrm{diss})=B'_\mathrm{1pc}\left(\frac{R'_\mathrm{diss}}{1\,\mathrm{pc}}\right)^{-1} \approx 9.0^{+120.0}_{-8.0}\,\mathrm{G}$. Our values of magnetic field strengths lie between the lower bound of this estimate and the best-fit value. These are consistent with the fact that in our case, the jet departs from equipartition, with $u'_\mathrm{particles}/u'_\mathrm{B} \gg 1$ with $u'_\mathrm{particles} = u'_\mathrm{p} + u'_\mathrm{e}$.

The disk is found to extend close to the black hole, $R_\mathrm{disk,in} \sim (4.5-5) \, R_\mathrm{g}$. The accretion disk and BLR, providing external photons to the emission region to interact with, have properties that stay overall constant throughout the epochs. This is due to the fact that external fields play a subdominant role in the emission compared to the internal photon fields, because the Doppler boosting into the jet frame is not important with the value of the Doppler factor that we find. Hence, the impact of the related parameters (i.e., $T_\mathrm{torus},\,r_\mathrm{diss}$) is negligible on the SED, leaving these two parameters essentially unconstrained for this set of solutions, when varying them independently.  

The primary electron populations injected in the emission region for each epoch have a spectral index $p_\mathrm{e} \sim 2.0$, typically expected for diffusive acceleration \citep{drury_introduction_1983} at strong shocks. Somewhat steeper values $p_\mathrm{e} \sim 2.2$ are expected from relativistic shocks in a weakly magnetized regime \citep[see, e.g.,][]{sironi_maximum_2013}. Although constraining a shock model is beyond the scope of this work, the energy-density ratios $u'_\mathrm{particles}/u'_\mathrm{B} \gg 1$ found in the solutions here hint qualitatively towards a weakly magnetized environment, supporting efficient shock acceleration \citep{sironi_particle_2011,sironi_maximum_2013,sironi_relativistic_2015}.

The primary protons are injected with a fixed index $p_\mathrm{p} = p_\mathrm{e}$ and a fixed minimum Lorentz factor $\gamma'_\mathrm{p,min} = 1.0$, but their injected luminosity is found to vary from $1 \times 10^{46}$ to $1\times10^{47} \, \mathrm{erg} \, \mathrm{s}^{-1}$, while the maximum energy at which they are injected is $E'_{\rm p, max} \sim 4-9 \, \mathrm{PeV}$. In terms of the contribution to the jet power, we can compare these injected physical luminosities $\Gamma_\mathrm{j}^2L'_\mathrm{p}/2$ to the Eddington luminosity $L_\mathrm{Edd} \approx 1.26 \times 10^{38}\, (M_\mathrm{BH}/M_\odot)\, \mathrm{erg}\,\mathrm{s}^{-1} \approx 8.75\times10^{46}\, \mathrm{erg}\,\mathrm{s}^{-1} $. We exceed the Eddington luminosity by a factor $2$ to $20$.

Due to the sub-dominance of the hadronic components, the SEDs of T4 and T5 are compatible with a proton injection luminosity $L'_\mathrm{p}=0$, however, in the other epochs, the contribution to the gamma-ray luminosity added by the hadronic component is required in order to explain the light curve (Figure \ref{fig:J0808_LC}). In T4, the neutrino prediction is below the axis limit of the plot. The primary populations injected in this epoch are strongly constrained by the gamma-ray data. In order to reproduce the cutoff in the GeV in the SED, the minimum Lorentz factor of the injected electrons is pushed to $\gamma'_\mathrm{e,min} = 7\times 10^3$, which in turns subtancially diminishes the radio and X-ray emission, compared to T1 for example, that has similar parameters but a less constraining gamma-ray spectrum.
The shaded areas in Fig. \ref{fig:J0808_SED_blob} demonstrate the constraints that a good multi-wavelength coverage in the SED has on the proton injection luminosity. When we vary $L'_\mathrm{p}$ even in T2 and T6 where the gamma-ray flux low and only comprise upper limits, the X-ray data are importantly constraining the range of fitting values, while in T1, T3, and T4, the constraint arises from the gamma-ray flux above $\mathrm{GeV}$ energies. 


\subsection{Neutrino expectations}

Over the six time intervals, the total muon neutrino luminosity varies from $L_{<\nu_\mu + \bar{\nu}_\mu>} \approx \mathbf {4\times 10^{43}} \, \mathrm{erg}\,\mathrm{s}^{-1}$ to $L_{<\nu_\mu + \bar{\nu}_\mu>} \approx 4 \times 10^{45} \, \mathrm{erg}\,\mathrm{s}^{-1}$. The highest predicted neutrino flux is obtained in epoch T3, which is contemporaneous to when the high-energy neutrino event \icevent\ was detected. In this epoch, the protons are injected with the largest luminosity, although there is no X-ray data to constrain the model further. Our models are predicting neutrinos of $\sim\ 195-820\,\rm TeV$ energy, which is right above the energy range where the IceCube detector is the most sensitive, at the source's declination (see the IceCube sensitivity curve in Fig. \ref{fig:J0808_SED_blob}). 

For each epoch, we compute the predicted number of neutrinos observable by the IceCube observatory, using the publicly available effective area of the detector\footnote{available at \url{https://IceCube.wisc.edu/data-releases/2021/01/all-sky-point-source-IceCube-data-years-2008-2018/}}, at the position of the source ($A_\mathrm{eff}(E,\delta)$), where $\delta$ is the declination of the source:
\begin{eqnarray*}
    N_{<\nu_\mu + \bar{\nu}_\mu>} = \int_{0}^{\infty} \int_{t_\mathrm{start}}^{t_\mathrm{stop}} dE \, dt \, A_\mathrm{eff}(E,\delta) \, \Phi_{<\nu_\mu + \bar{\nu}_\mu>}(E,t),
\end{eqnarray*}
where $\Phi_{<\nu_\mu + \bar{\nu}_\mu>}(E,t)$ is the sum of the muon neutrino and anti-neutrino flux in units of neutrinos per unit energy per unit of time per unit of surface, predicted by the models. The number of neutrinos predicted per epoch is reported in the last row of Table \ref{tab:state_parameters}.

Similarly to the flux, the highest predicted neutrino rate is obtained in epoch T3. Assuming that the predicted number of neutrinos in each epoch follows a Poisson distribution, the $3\sigma$ confidence interval is reported along with the predicted number of neutrino per epoch in Table \ref{tab:state_parameters}, as an asymmetrical uncertainty. For epoch T3, this corresponds to $[0-2]$ neutrinos, which is compatible with the detection of one neutrino event over these $4.9\,\mathrm{years}$.
The corresponding probability to detect at least one neutrino during this epoch, still assuming Poisson statistics, is $P(N_\mathrm{det} \geq 1) = 1-\exp(-N_{<\nu_\mu+\bar{\nu}_\mu>}) \approx 9.6\%$.

Cumulating all six epochs $\sum N_{<\nu_\mu + \bar{\nu}_\mu>} $, we predict a total number of $0.17$ neutrinos observable by IceCube in $16.4\,\mathrm{years}$ for the best-fit case, corresponding to a probability of detecting at least one neutrino $P(N_\mathrm{det} \geq 1)  \approx 16$

\section{Discussion}
\label{sec:discussion}

\target\ has been studied by several authors in the context of population studies \citep{paliya_central_2021, rodrigues_leptohadronic_2024, finke_properties_2019, hovatta_doppler_2009}, particularly related to estimating the Doppler factor of the core, and possible core shifts. In \cite{finke_properties_2019} the modeling reports $ 1.1 \leq \Gamma_\mathrm{j} \leq 11.9 $, $ 1.2 \leq \delta_\mathrm{j} \leq 15.1 $ and $0.1 \leq B_\mathrm{1 pc} \leq 12.9$, while \cite{hovatta_doppler_2009} report a Doppler factor of $\sim 35$, based on the light curve variability timescale. The Lorentz factor value obtained through our fitting method lies within the estimates \cite{finke_properties_2019}, which is consistent with the lack of variability that we observe in the light curve.
The predicted SED is also consistent with the classification of \target\ as a low-synchrotron-peak blazar, i.e., the synchrotron radiation from the primary electrons peaks at frequency $\nu_\mathrm{syn} \lesssim 10^{14} \,  \mathrm{Hz}$ while the high-energy hump of the SED is peaking around $\nu_\mathrm{IC} \gtrsim 10^{20} \,  \mathrm{Hz}$ when modeled by a purely leptonic model in \cite{paliya_central_2021}.

In \cite{paliya_central_2021} and \cite{rodrigues_leptohadronic_2024}, this source is found to be consistent with purely leptonic models. This is compatible with our findings in epochs T4 and T5, given the sub-dominance of the proton-initiated components, as visible in Fig. \ref{fig:J0808_SED_blob}, where we estimate the range of possible proton luminosity to be compatible with $0$. 
However, the previous studies were focusing only on the average SED of \target, which we can see in Fig. \ref{fig:J0808_SED_blob} with the gray data points. The way we model the SED is different from their approach, because we use a time-dependent assumption. As can be seen in Fig. \ref{fig:J0808_SED_blob}, when the source exhibits a lower gamma-ray flux (T2 and T6), the SED differs significantly from the average, and the relative contribution of the hadronic components is more prominent than during the gamma-ray-enhanced states.

The jet power predicted by our model is larger than the Eddington luminosity for this source, indicating that the kinetic output is larger than what can be powered by accretion. This is a common conclusion of lepto-hadronic models for neutrino-emitter-candidate blazars \citep[e.g.][]{keivani_multimessenger_2018,gao_modelling_2019,cerruti_leptohadronic_2019,liodakis_proton_2020}, mainly due to the contribution of relativistic protons in our case. The Eddington luminosity is not in itself a fundamental limit on the jet power. Another quantity one can compare the jet power to is the power of the Blandford-Znajek process \citep{blandford_electromagnetic_1977}. The predicted jet power we find is within the estimates of \cite{liodakis_proton_2020} $P^\mathrm{jet}_\mathrm{BZ} = 10^{47.0\pm1.38} \, \mathrm{erg/s}$. Nevertheless, even if jets carrying powers of the order of $\sim \dot{M}c^2$ (i.e., $P_\mathrm{jet} \approx1.5\times 10^{47} \, \mathrm{erg/s}$ in our case) are not uncommon \citep{tchekhovskoy_efficient_2011,zamaninasab_dynamically_2014, ghisellini_power_2014}, the typical efficiency is assumed to be of the order of $\sim0.1\dot{M}c^2$ \citep[e.g][]{mckinney_total_2005}.
A way to ease the tension with the jet power would be if protons are accelerated with a harder spectral index $p_\mathrm{p} \rightarrow 1$ \citep[through magnetic reconnection, studies show that the index can be as hard as 1.2 in the case of large magnetization, see e.g][]{guo_formation_2014, sironi_relativistic_2014, werner_extent_2016}. With the current energy-density ratios between magnetic field and protons, we find, however, that acceleration through magnetic reconnection would be inefficient \citep[see, e.g.][for a review]{guo_magnetic_2024}, and hence, in order to reach such hard power-law injection, the particles would need to be accelerated outside the blob, in highly magnetized regions. The photo-hadronic interactions are dominated by the high-energy protons, so a harder spectrum will lead to a larger fraction of those high-energy protons compared to the low-energy ones. However, the neutrino predictions would not drastically change, given the constraints on the subdominance of the hadronic contribution to the total SED.

\begin{figure*}
    \centering
    \includegraphics[width=1\linewidth]{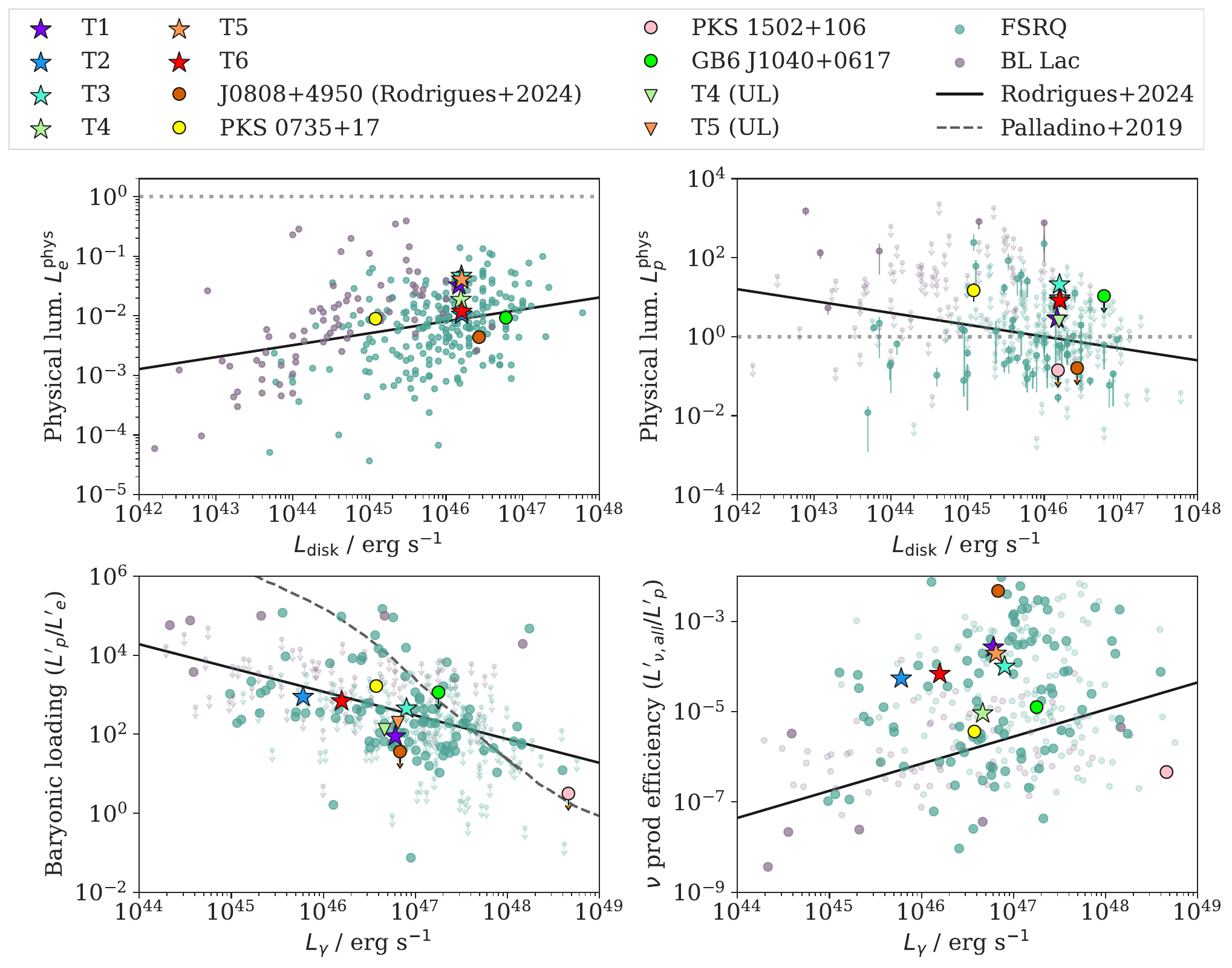}
    \caption{Comparison with the physical quantities from the 324-blazar sample of \cite{rodrigues_leptohadronic_2024}. The cyan blue points represent the FSRQs of the sample, while the purple points show the BL Lacs. The best-fit values are represented with the filled circles, the empty circles in the upper right and lower panels represent the sources for which the modeling is compatible with a purely leptonic model (i.e., the lowest possible proton injection luminosity is zero) while the plain triangles show the sources for which the best-fit is purely leptonic and in this case we use the maximum proton luminosity to compute the physical luminosity, the baryonic loading and the neutrino production efficiency, so these points are considered upper limits.
    The quantities obtained by modeling \target\ in \cite{rodrigues_leptohadronic_2024} are shown in red, while the ones we derive are shown with the stars or with a downward triangle when the proton luminosity for the epoch is compatible with $0$. We also indicate the results of this previous study for three IceCube neutrino candidate sources, \pksseven, \pksfifteen, and \gbneutrino.
    The solid lines show the relations derived in \cite{rodrigues_leptohadronic_2024}, the dotted gray lines show the equality relation, and the dashed line gives the relation derived in \cite{palladino_interpretation_2019}.
    \textit{Upper panels}: physical luminosity $\Gamma_\mathrm{j}^2L'_\mathrm{e/p}/2$ as a function of the disk luminosity \citep[obtained from][for the blazar sample]{paliya_general_2017}. \textit{Lower left:} baryonic loading ($L'_\mathrm{p}/L'_\mathrm{e}$) as a function of the gamma-ray luminosity. For the blazar sample, we use the Fermi-LAT flux reported in \cite{rodrigues_leptohadronic_2024}, and for this work, we integrate the modeled SEDs between $100 \, \mathrm{MeV}$ and $100 \, \mathrm{GeV}$. \textit{Lower right:} All-flavor neutrino production efficiency $L'_{\nu,\mathrm{all}}/L'_\mathrm{p}$ as a function of the gamma-ray luminosity.}
    \label{fig:comparison_panels}
\end{figure*}

\target\ shares common characteristics with other FSRQs. This can be seen in Fig. \ref{fig:comparison_panels}, where we show the source's physical luminosities ($L^\mathrm{phys}_\mathrm{e/p} = \Gamma_\mathrm{j}^2L'_\mathrm{e/p}/2$) in Eddington units (i.e., normalized to the Eddington luminosity of each source), baryonic loading ($L'_\mathrm{p}/L'_\mathrm{e}$) and neutrino production efficiency $L'_{\nu,\mathrm{all-flavor}}/L'_\mathrm{p}$ as a function of disk and gamma-ray luminosities. The baryonic loading is used to compare the power injected into protons, which are needed to produce neutrinos, and the power injected into electrons, which are the primary population contributing the most to the SED. However, a large baryonic loading does not necessarily imply efficient neutrino production, which is why we also compare the neutrino production efficiency to the gamma-ray luminosity. We compare our results with the study of 324 blazars proposed by \cite{rodrigues_leptohadronic_2024}. The FSRQ are represented by the cyan points, while BL Lac objects are shown in purple. The red cross shows the results of the modeling of \target\ in the same paper, while the stars show the quantities derived in this present work. The relations between quantities derived in \cite{rodrigues_leptohadronic_2024} are shown with the black solid lines. 

Our models predict electrons and protons luminosities characteristic of the powerful accretors, as shown in the top right and left panels of Fig. \ref{fig:comparison_panels}, where \target\ is located where the largest disk luminosities are found.
In the lower left panel of Fig. \ref{fig:comparison_panels}, we show the baryonic loading and the curve derived by \cite{palladino_interpretation_2019} that shows the evolution of the baryonic loading with respect to the gamma-ray luminosity in order not to violate the IceCube stacking limit. The predicted baryonic loading in \target\ is consistent with the fact that no significant excess in neutrino emission has been observed towards the blazar population. Although the baryonic loading is low for \target\ the neutrino production efficiency (that is defined as the ratio of the all-flavor comoving neutrino luminosity over the comoving proton luminosity) is larger than for other FSRQs, which is the main peculiarity of \target\ compared to the global FSRQ population. In Fig. \ref{fig:comparison_panels}, lower right panel, the source lies above the trend derived in \cite{rodrigues_leptohadronic_2024}. As a comparison, we show three candidate neutrino sources, \pksseven, \pksfifteen\  and \gbneutrino\ in yellow, pink, and lime, respectively. In comparison with \pksseven\  and \gbneutrino, which have gamma-ray luminosities of the same order as \target\, we predict a higher neutrino luminosity, of the order of $1$ to $40$ times larger. However, this prediction relies only on the total integrated neutrino luminosity and not on the neutrino energy, which is a feature to assess whether a source can be an IceCube neutrino candidate. 
Our prediction lies below the efficiency given by \cite{rodrigues_leptohadronic_2024}, but this value is given as an upper limit on the proton and neutrino luminosities, as they found the source to be compatible with a purely leptonic model on average.

\begin{figure*}
    \centering
    \includegraphics[width=\textwidth]{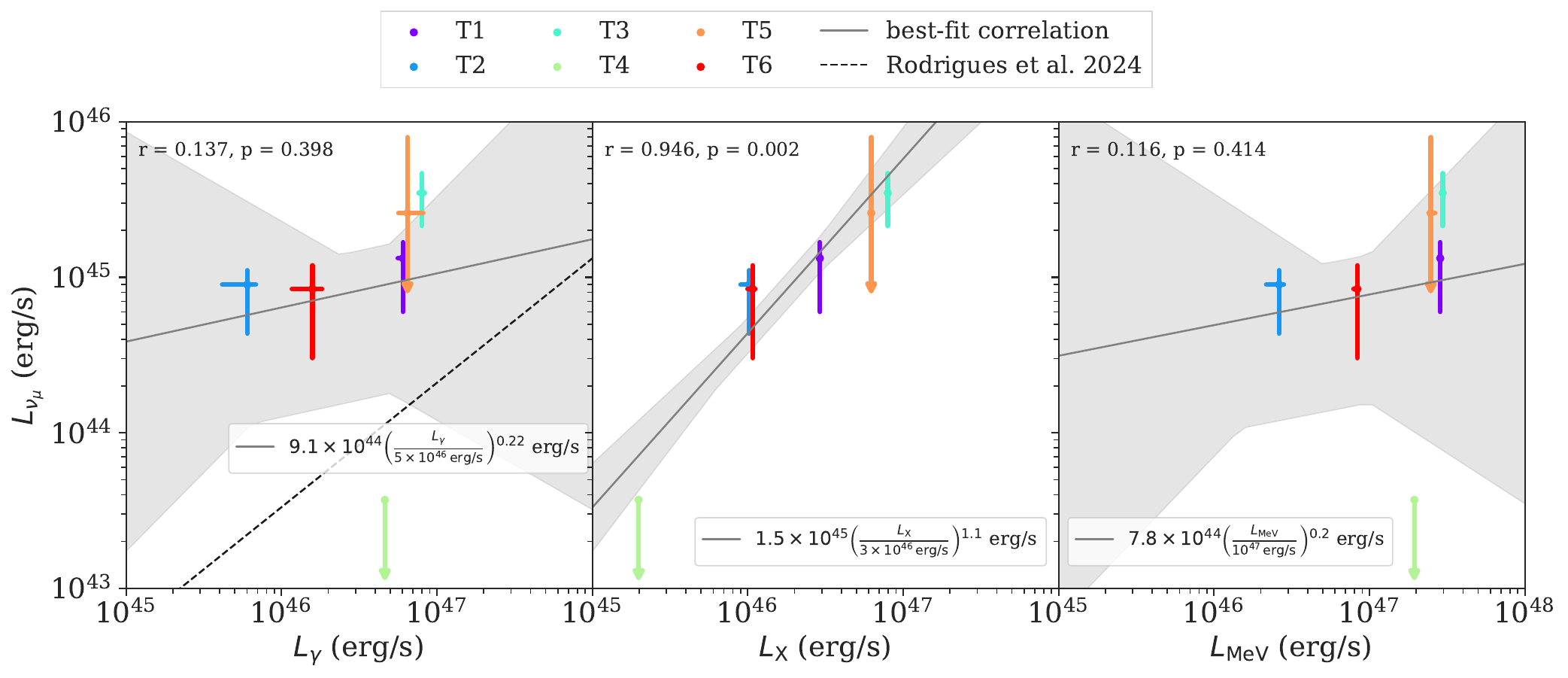}
    \caption{Comparison between the predicted neutrino luminosity and the gamma-ray ($0.1-100\,\rm GeV$), X-ray ($0.3-10\,\rm keV$), and MeV ($0.1-100\,\rm MeV$) luminosities obtained for the best-fit solutions of \ref{tab:state_parameters}, from left to right. The linear best-fit results are indicated in each panel. The grey shaded areas represent the $1\sigma$-uncertainty band on the slope and the intercept of the lines. We report the Pearson correlation coefficient and the corresponding p-value on the bottom left of each panel. The best-fit correlation found in \cite{rodrigues_leptohadronic_2024} is shown with the dashed black line in the left panel.} 
    \label{fig:luminosities}
\end{figure*}

We also compare the predicted neutrino luminosity with three energy-band luminosities from our best-fit models for the different epochs in Fig. \ref{fig:luminosities}. Namely, we investigate whether there is a correlation with either of the gamma-ray (integrated between $0.1\,\rm GeV$ and $100\,\rm GeV$), the X-ray (integrated between $0.3\,\rm keV$ and $10\,\rm keV$) or the MeV (integrated between $0.1\,\rm MeV$ and $100\,\rm MeV$) luminosity. Each luminosity is calculated from the SED prediction for each epoch. We perform a least-square regression to find the correlation $L_{\nu_\mu} = b\,\left(\frac{L_\mathrm{i}}{L_\mathrm{i, scale}}\right)^a$ where $L_i$ is one of the three luminosities mentioned previously, and $L_\mathrm{i, scale}$ is the scale of this given quantity over the six epochs. The results of the fits are given in each panel of Fig. \ref{fig:luminosities}. We report the best-fit values and their uncertainty in Table \ref{tab:luminosities}. 
\begin{table}
    \centering
    \begin{adjustbox}{width=\linewidth}

    \begin{tabular}{|c|c|c|c|c|}
        \hline
         Relation & $a$  & $ b / 10^{45}$ & $L_\mathrm{scale}$ / $10^{46}$\\
         \hline
         $\log{L_{\nu_\mu}} - \log{L_\gamma}$         & $0.22 \pm 0.79$  & $0.92 \pm 0.73$ & $5.00$\\
         $\log{L_{\nu_\mu}} - \log{L_\mathrm{X}}$     & $1.12 \pm 0.19$  & $1.50 \pm 0.39$ & $3.00$\\
         $\log{L_{\nu_\mu}} - \log{L_\mathrm{MeV}}$   & $0.20 \pm 0.85$  & $0.78 \pm 0.62$ & $10.0$ \\
         \hline
    \end{tabular}
    
    \end{adjustbox}

    \caption{Best-fit results of the correlation between the neutrino luminosities and the other luminosities, presented in Fig. \ref{fig:luminosities}.}
    \label{tab:luminosities}

\end{table}

We find that, if a hadronic component is present, as in the best-fit solutions we present here, the predicted muon- and anti-muon-neutrino luminosity correlates best with the X-ray luminosity, visible in the second panel of Fig. \ref{fig:luminosities}. 
The MeV band has a tendency to be even less correlated to the neutrino luminosity. Even if currently there are no constraints on the MeV data, and hence, depends strongly on the model prediction, this energy band is typically where the electromagnetic cascade induced by the presence of protons emerges \citep[see, e.g][]{rodrigues_leptohadronic_2019, reimer_cascading_2019, petropoulou_multiepoch_2020}, as can be seen in epoch T2 in Figure \ref{fig:J0808_SED_blob}. For this reason we explored whether the MeV band could be an indicator of the neutrino production, even for subdominant hadronic components. We do not find a significant correlation, but the addition of MeV data with future instruments such as the Compton Spectrometer and Imager \citep[COSI,][]{tomsick_compton_2019, tomsick_compton_2021}, All-sky Medium Energy Gamma-ray Observatory eXplorer \citep[AMEGO-X,][]{caputo_allsky_2022a}, or enhanced ASTROGAM \citep[e-ASTROGAM,][]{deangelis_science_2018} would nonetheless be crucial in constraining the SED models of this source, given the strong emission we find in this band.
The neutrino-to-gamma-ray ratio $Y_{\nu\gamma} = L_{\nu_\mu+\bar{\nu}_\mu}/L_\gamma$ for the 6 epochs varies between $8\times 10^{-4}$ (in the epochs where the Compton dominance\footnote{the Compton-dominance parameter $\mathrm{CD}$ defined as the ratio between the peak of the inverse-Compton component over the peak of synchrotron flux from the primary electrons \citep[see, e.g][]{fossati_unifying_1998}, and is reported in the Table \ref{tab:state_parameters}} is larger than unity) and $1.5\times 10^{-1}$ (in T2, where the Compton dominance is the lowest). In the time interval T3, in which the neutrino flux is predicted with the largest value, $Y_{\nu\gamma} \approx 4.3\times 10^{-2}$. 
These values are compatible with the $5-10\%$ ratio found in several studies \citep[e.g][]{aartsen_contribution_2017, wang_cumulative_2016, padovani_simplified_2015, capel_assessing_2022} assessing the contribution of gamma-ray blazars to the diffuse neutrino flux detected by IceCube.
In comparison, the trend for the total population of FSRQs and BL Lacs studied in \cite{rodrigues_leptohadronic_2024} is shown as the dashed black line. We find a trend $L_{\nu_\mu} \propto L_\gamma^{0.22\pm0.79}$, which is compatible within the uncertainty to the population trend, at the level of an individual source. The correlation between the X-ray luminosity and the muon-neutrino luminosity has a smaller uncertainty than the gamma-ray correlation and a larger Pearson r-coefficient. This hints that the X-ray variations of \target\ are a better indicator of the neutrino activity over time.

\section{Conclusions}
\label{sec:conclusion}
We have studied the neutrino-candidate flat-spectrum radio quasar \target\ in the context of six contemporaneous, multi-wavelength epochs. We used the numerical, semi-analytical software AM$^3$ in order to derive time-averaged lepto-hadronic SEDs for each epoch. We modeled the emission region as a sphere moving relativistically, and took into account the influence of external photons emitted by an accretion disk and reprocessed by a BLR and a dusty torus, as the emission region travels along the jet axis. 

We find solutions with compact emission regions and moderate bulk Lorentz factor, where the changes in the primary particles are the main variation over the epochs, with only a subdominant hadronic component and a moderate neutrino flux peaking below $\mathrm{PeV}$ energies. The contribution of the hadronic cascade is visible in the high-energy gamma rays, above GeV energies, while the rest of the SED is dominated by the leptonic components.
The neutrino flux predicted in this work stays below the IceCube detector sensitivity, with a predicted rate ranging from $1\times 10^{-3}$ to $1\times 10^{-1}$ neutrinos per epoch. With the future upgrades of the IceCube observatory, namely IceCube-Gen2 \citep{aartsen_icecubegen2_2021}, the source may be at reach of the more sensitive detector. The sensitivity is currently expected to be better by a factor of 5. The highest rate is obtained in the epoch T3, contemporaneous with the IceCube high-energy event \icevent. Assuming Poisson statistics on the mean number of neutrinos detectable, the $1\times 10^{-1}$ neutrino number predicted during the epoch T3 is compatible with the detection of the high-energy neutrino \icevent.
Our model shows a stronger correlation between the predicted neutrino luminosity and the X-ray luminosity, rather than with the gamma-ray luminosity in the case of the best solutions including a hadronic component.
In the context of modeling neutrino candidates, this may suggest that monitoring the X-ray variability of a source is a better proxy of neutrino emission in a given source than the gamma-ray variability. 

\begin{acknowledgements}
We thank the anonymous referee for the insightful comments on the manuscript.
This work was supported by the European Research Council, ERC Starting grant \texttt{MessMapp}, S.B. Principal Investigator, under contract no. 949555. 
This research has made use of data and/or software provided by the High Energy Astrophysics Science Archive Research Center (HEASARC), which is a service of the Astrophysics Science Division at NASA/GSFC.
This research furthermore has made use of the NASA/IPAC Extragalactic Database, which is funded by the National Aeronautics and Space Administration and operated by the California Institute of Technology.
Part of this work is based on archival data, software or online services provided by the Space Science Data Center - ASI.
We thank the developers of AM3 for providing the framework used to interpret the data used for this work.
The \textit{Fermi}-LAT Collaboration acknowledges support for LAT development, operation, and data analysis from NASA and DOE (United States), CEA/Irfu and IN2P3/CNRS (France), ASI and INFN (Italy), MEXT, KEK, and JAXA (Japan), and the K.A.~Wallenberg Foundation, the Swedish Research Council, and the National Space Board (Sweden). Science analysis support in the operations phase from INAF (Italy) and CNES (France) is also gratefully acknowledged. This work was performed in part under DOE Contract DE-AC02-76SF00515.
We acknowledge the use of public data from the Neil Gehrels Swift Observatory data archive.
This work presents results from the European Space Agency (ESA) space mission Gaia. Gaia data are being processed by the Gaia Data Processing and Analysis Consortium (DPAC). Funding for the DPAC is provided by national institutions, in particular the institutions participating in the Gaia MultiLateral Agreement (MLA). The Gaia mission website is \url{https://www.cosmos.esa.int/gaia}. The Gaia archive website is \url{https://archives.esac.esa.int/gaia}.
This work is based on observations obtained with the Samuel Oschin Telescope 48-inch and the 60-inch Telescope at the Palomar Observatory as part of the Zwicky Transient Facility project. ZTF is supported by the National Science Foundation under Grant No. AST-2034437 and a collaboration including Caltech, IPAC, the Weizmann Institute for Science, the Oskar Klein Center at Stockholm University, the University of Maryland, Deutsches Elektronen-Synchrotron and Humboldt University, the TANGO Consortium of Taiwan, the University of Wisconsin at Milwaukee, Trinity College Dublin, Lawrence Livermore National Laboratories, and IN2P3, France. Operations are conducted by COO, IPAC, and UW.\\
This research has made use of the RATAN-600 data, provided by the Special Astrophysical Observatory of the Russian Academy of Sciences (SAO RAS).
We use radio data from the CATS database (\url{www.sao.ru/cats/}), operated at SAO RAS, Russia \citep{verkhodanov_current_2007}.
This research has made use of data from the OVRO 40-m monitoring program. This monitoring program has been supported by NSF grants AST-0808050 and AST-1109911, and is currently supported by NSF grants AST-2407603 and AST-2407604. It has also been supported by NASA grants NNX08AW31G, NNX11A043G, and NNX14AQ89G.
\end{acknowledgements}

\bibliography{zotero}{}
\bibliographystyle{aasjournal}

\begin{appendix}
\counterwithin{figure}{section}
\section{\textit{Fermi}-LAT data analysis insights and results}\label{Appendix: Fermi}
\begin{table*}[h]
    \centering
    \begin{adjustbox}{width=0.8\textwidth}
    \begin{tabular}{|c|c|c|c|c|c|}
         \hline
         Interval&Time&$F_{1\,\mathrm{GeV}-1000\,\mathrm{GeV}}$&$\alpha$&$\beta$&TS\\
           &&$\times10^{-10}$&&&\\
            &MJD&$\mathrm{ph}\,\mathrm{cm}^{-2}\,\mathrm{s}^{-1}$&&&\\
            \hline
             T1& $54682-55047$& $1.34\pm0.60$& $2.91\pm0.37$& $0.38\pm0.48$&21.09\\
             T2& $55047-55412$& $4.43$& -& -&$3.80$\\
             T3& $55412-57237$& $2.76\pm0.39$& $2.93\pm0.09$& $-0.07\pm0.05$&177.47\\
             T4& $57237-58687$& $0.67\pm0.25$& $3.37\pm 0.28$& $0.03\pm0.18$&52.26\\
             T5& $58687-60157$& $2.43\pm0.62$& $2.71\pm0.16$& $-0.04\pm0.06$&65.72\\
             T6& $60157-60675$& $1.71$& -& -&1.75\\
 \hline
    \end{tabular}
    \end{adjustbox}
    \caption{\textit{Fermi}-LAT analysis results for the individual time intervals analyzed for \targetgamma. The analysis procedure is summarized in Appendix \ref{Appendix: Fermi} and the gamma-ray SEDs created from the analysis steps are used for the modelling shown in Figure \ref{fig:J0808_SED_blob}. If the source is not significantly detected for the time interval, the 95\%  flux upper limit is reported.}
    \label{tab:fermi}
\end{table*}
For the analysis of \targetgamma\ we extracted photons of \texttt{Pass 8 SOURCE}
and applied a zenith angle cut of $90^\circ$ to limit the impact of high energy (HE) photons produced in Earth's upper atmosphere. In a next step fermipy performs a binning both in the spatial and the energy distributions. For this we adopted a pixel size of $0.08\deg$ and 12 energy bins per decade. We furthermore took into account energy dispersion for point sources, but excluding the isotropic diffuse component to avoid modeling inconsistencies with older diffuse templates as suggested in the fermipy documentation \footnote{More details at \url{https://fermipy.readthedocs.io/en/latest/}}. Beyond the $10^\circ\times10^\circ$ photon information, the pre-fit ROI model furthermore includes the source models of all known sources of the 4FGL-DR4 catalog \citep{ballet_fermi_2023,abdollahi_incremental_2022} located within $15^\circ$ from the ROI center, as well as the corresponding galactic and isotropic diffuse emission (\texttt{gll\_iem\_v07.fits} and \texttt{iso\_P8R3\_SOURCE\_V3\_v1.txt}\footnote{More details at \url{https://fermi.gsfc.nasa.gov/ssc/data/access/lat/BackgroundModels.html}}) \citep{atwood_pass_2013,bruel_fermilat_2018}. The analysis is performed in 2 steps. First we analyzed the gamma-ray properties of the source for the full-time interval performing the following steps:
\begin{enumerate}
    \item Using the configuration stated before, we run an initial optimization (using fermipy's \texttt{optimize()} function) to possibly account for changes in the source models that are taken from the 14-year catalog while we now analyze $\sim16$ years of data. The function uses an iterative fitting procedure so that ideally all the source parameters in the ROI are close to their global likelihood maxima. 
    \item We then search for the existence of possible new significant gamma-ray emission using fermipy's \texttt{find\_sources()} function, assuming a point source with its spectral shape following a power law, an excess of at least $\sqrt{TS}=5$ and a minimum distance to its nearest neighbor of at least $r_{min}=0.5^{\circ}$.
    \item With fermipy's \texttt{localize()} function we then scan the likelihood surface of a local patch around the source of interest to refine its position in the ROI. 
    \item The final best fit model (with fermipy's \texttt{fit()} function) is then computed by freeing the normalizations and spectral parameters for \targetgamma\ and all sources within a radius of $r_{min}=3^{\circ}$, as well as the normalization for the diffuse fluxes and all sources with $TS\geq10$. We allow the fitting function to retry $3$ times to converge to a good fit quality.
    \item Finally we compute the gamma-ray SED and lightcurve using fermipy's \texttt{sed()} and \texttt{lightcurve()} functions.
\end{enumerate}
In step $2$ we find the new point source \texttt{PS J0738.8+5336}. We obtain a flux of $F_{1\,\mathrm{GeV}-1000\,\mathrm{GeV}}=(1.14\pm0.23)\times 10^{-10}\,\mathrm{ph}\,\mathrm{cm}^{-2}\,\mathrm{s}^{-1}$ and an index of $\mathrm{Index}=-2.38\pm0.14$ which, within its errors, is consistent with the values for \texttt{FL16Y J0738.7+5336} that was added to the recent 16-year source list of \textit{Fermi}-LAT \citep[FL16Y;][]{ballet_fermilat_2026}. There is no known multi-wavelength counterpart associated to this source yet.
The relocalization of \targetgamma\ resulted in the new coordinates $\mathrm{RA} = 122.17^\circ\pm0.41^\circ$ and $\mathrm{DEC}= 49.79^\circ\pm0.41^\circ$, which within its bounds is consistent with the coordinates reported in FL16Y for \texttt{FL16Y J0808.5+4947}, associated with our source. \targetgamma\ is reported to follow a log-parabolic spectral shape. For this we obtain the following best fit values: $F_{1\,\mathrm{GeV}-1000\,\mathrm{GeV}}=(1.94\pm0.25)\times 10^{-10}\,\mathrm{ph}\,\mathrm{cm}^{-2}\,\mathrm{s}^{-1}$, $\alpha=2.82\pm0.09$ and $\beta=0.01\pm0.06$. The spectral index $\alpha$ is steeper than the value reported in FL16Y ($\alpha_{\mathrm{ref}}=2.70\pm0.09$) and the curvature value $\beta$ contributed less to the spectral shape of the best fit model ($\beta_{\mathrm{ref}}=0.12\pm0.07$) however, the differences are mild and overlap in their uncertainties. They might arise from differences in the analysis procedure and differences in the energy range taken into account for the catalog.\\
For the analysis of the shorter time intervals we repeated the same analysis procedure, however we included the new point source \texttt{PS J0738.8+5336} to the ROI of each analysis and fixed \targetgamma\ to the position we obtained from the localization, to ensure a well constrained model and consistency throughout the analysis steps. No additional point sources are found for the shorter integration times. A summary of the results is displayed in Table \ref{tab:fermi}. 
\end{appendix}
\end{document}